\documentclass[final,twocolumn]{elsarticle}

\usepackage{amssymb}
\usepackage{amsmath}
\usepackage{graphicx}% Include figure files
\usepackage{dcolumn}% Align table columns on decimal point
\usepackage{bm}% bold math
\usepackage{xcolor}
\usepackage{multirow}
\usepackage{makecell}
\usepackage{placeins}
\usepackage{soul}
\usepackage{lineno}
\usepackage{xurl}
\usepackage{hyperref}% add hypertext capabilities
\usepackage{ulem}

\setul{0.5ex}{0.3ex}
\setulcolor{red}

\journal{Journal of Materiomics}

\begin{document}

\begin{frontmatter}



\title{Magnetoelasticity -  magnetic structure interrelation -- tetragonal MnPt  system study} %% Article title


\author[VSB]{Jakub Šebesta} %% Author name
\ead{jakub.sebesta@vsb.cz}

\author[PAS_P]{Karol Synoradzki}%
%\affiliation{Institute of Molecular Physics, Polish Academy of Sciences, Smoluchowskiego 17, 60-179 Poznań, Poland}%

\author[UK]{Michal Vališka}%
%\affiliation{Charles University, Faculty of Mathematics and Physics
%Ke Karlovu 3, 121 16 Praha 2, Czech Republic}%

\author[UK]{Tetiana Haidamak}%
%\affiliation{Charles University, Faculty of Mathematics and Physics
%Ke Karlovu 3, 121 16 Praha 2, Czech Republic}%

\author[PAS_W]{Tamara J. Bednarchuk}%
%\affiliation{Institute of Low Temperature and Structure Research, Polish Academy of Sciences, Okólna 2, 50-422 Wrocław, Poland}%

\author[OV]{{Pablo Nieves}}%

\author[UK,VSB]{Dominik Legut}

\affiliation[VSB]{organization={IT4Innovations, VSB – Technical University of Ostrava},%Department and Organization
            addressline={17. listopadu 2172/15}, 
            city={Ostrava},
            postcode={708 00},
            country={Czech Republic}}

\affiliation[PAS_P]{organization={Institute of Molecular Physics, Polish Academy of Sciences},%Department and Organization
            addressline={Smoluchowskiego 17}, 
            city={Poznań},
            postcode={60-179},
            country={Poland}}

\affiliation[UK]{organization={Charles University, Faculty of Mathematics and Physics},%Department and Organization
            addressline={Ke Karlovu 3}, 
            city={Praha 2},
            postcode={121 16},
            country={Czech Republic}}

\affiliation[PAS_W]{organization={Institute of Low Temperature and Structure Research, Polish Academy of Sciences},%Department and Organization
            addressline={Okólna 2}, 
            city={Wrocław},
            postcode={50-422},
            country={Poland}}

\affiliation[OV]{organization={Departamento de Física, Universidad de Oviedo},%Department and Organization
            addressline={C. Leopoldo Calvo Sotelo, 18}, 
            city={Oviedo},
            postcode={33007},
            country={Spain}}

%% Abstract
\begin{abstract}
Magnetic materials represent an essential ingredient for the contemporary industry. Apart from common material parameters such as magnetocrystalline anisotropy, coercivity, or saturation magnetization,  magnetoelastic behavior is vital for applications serving in various devices, {\it e.g.}, in acoustic actuators, transducers, or sensors providing a desirable fast response and high efficiency with respect to applied magnetic field. 
Magnetoelastic properties have been studied  for ferromagnetic 3d elements, or especially in high symmetry systems containing rare-earth elements to achieve higher values. Since, unlike for rare earth Laves phases, in the transition metals or alloys, these effects are very weak. 
Here, in contrast, we analyze the 
magnetoelastic behavior of antiferromagnetic tetragonal system MnPt, {explaining the experimentally measured data based on the theoretical calculations and} discussing the influence of the magnetic structure. {Particularly, we inspect} the origin of magnetocrystalline anisotropy energy, as well as the size and source of the isotropic and anisotropic parts of magnetoelastic (magnetostriction) coefficients. 
\end{abstract}

% \includegraphics[width=\textwidth]{graphical_abstract.pdf}

% \begin{highlights}
% \item Magnetoelastic properties of MnPt
% \item Theoretical explanation the of experimentally measured magnetostriction
% \item Modification of the magnetoelastic properties with respect the magnetic structure
% \item Origin of the magnetic structure dependence in the magnetoelastic behavior
% \end{highlights}

%% Keywords

\end{frontmatter}



%%%%%%%%%%%%%%%%%%%%%%%%%%%%%%%%%%%%%%%%%%%%%%%%%%%%%%%%%%%%%%%%%%%%%%%%%%%%%%%%%%%
%  Introduction 
%%%%%%%%%%%%%%%%%%%%%%%%%%%%%%%%%%%%%%%%%%%%%%%%%%%%%%%%%%%%%%%%%%%%%%%%%%%%%%%%%%%

\section{Introduction}
%\section{Introduction}

The mechanical properties of the material are important for practical applications.  However, for magnetic materials, high complexity emerges due to the coupling of magnetic and elastic properties, giving rise to the so-called magnetoelastic behavior~\cite{chikazumi2009physics}. There exist various kinds of magnetoelastic effects, e.g. {volume and axial magnetostriction} or Wiedemann effect, bringing many useful applications. Thereby, magnetoelastic  effects can take place in force sensors, low-frequency weak-field detectors, fast pulse generators,  rotational and linear motors, or magneto-acoustically driven low-energy demanding electronics~\cite{Ekreem_jmpt_r07,Spetzler_scirep_r21,Calkins_jimss_r07,Bienkovsky_SensAct_r04,Kuszewski_iop_r18}. {Magnetoelasticity arises from the mutual relation between deformation of crystal lattice and the magnetocrystalline anisotropy.}  In the present work, we focus {by ab-initio calculations} on the {interrelation between magnetic structure and magnetic properties of MnPt system}. Magnetoelastic behavior, which stems from an interplay between the magnetic and elastic properties, substantially differs among the magnetic compounds. The MnPt might bear interesting properties as it belongs to the family of Pt-based tetragonal systems~\cite{Nieves_sss_r25_seconorder,Das_JoPD_r25_FePt}, where a large magnetocrystalline anisotropy occurs~\cite{Ravindran_prb_r01}. It comes from the interplay of Pt strong spin-orbit coupling (SOC) with  spin  moments of the 3$d$-elements~\cite{Soares_prb85_r12,Shick_prb78_r08,QURATULAIN_jmmm_r18,Hama_jpcm_r07}.  
At room temperature, MnPt forms the L1$_0$ phase~\cite{Kubota_APL_r07,Kang_prb_r23,Hama_jpcm_r07,pal_jap_r68}. It was found that such an Mn-based two-component alloy possesses surprisingly high N\'{e}el temperature ($T_{N}> $900~K)~\cite{Kren_prb_r68,Hama_jpcm_r07,Kang_prb_r23,pal_jap_r68}. The antiferromagnetic ordering of Mn-based systems arises from around half-filled Mn $d$-states~\cite{Kubota_APL_r07} and the magnetic structure of MnPt is characterized by [001] N\'{e}el vector~\cite{Kang_prb_r23}. 
MnPt magnetic properties are useful in tunnel magnetoresistance devices~\cite{Schwickert_jap89_r01,Childress_jap89_r01} or for spin-valve structures with giant magnetoresistance
~\cite{Saito_jap85_r99,Khapikov_jap93_r03,Kubota_APL_r07}. Besides, the MnPt $a$ lattice parameter is close to the Fe or MgO ones, which makes MnPt interesting for creating interfaces with those constituents~\cite{AISSAT_jmmm_r22}.

In this work, we investigate the impact of different magnetic structures on the magnetoelastic behavior based on ab-initio calculations and  explain the experimental behavior of a prepared MnPt sample employing ab-initio results and atomistic spin simulations.
{Unlike recent papers, it  includes a description of antiferromagnetic systems.}
The calculated results  reveals a  {substantial modifications} of the magnetoelasticity with respect to the considered magnetic structures. We analyze  the origin of the observed changes  investigating modifications in magnetocrystalline anisotropy and charge density.
% 

%%%%%%%%%%%%%%%%%%%%%%%%%%%%%%%%%%%%%%%%%%%%%%%%%%%%%%%%%%%%%%%%%%%%%%%%%%%%%%%%%%%
%  Methods 
%%%%%%%%%%%%%%%%%%%%%%%%%%%%%%%%%%%%%%%%%%%%%%%%%%%%%%%%%%%%%%%%%%%%%%%%%%%%%%%%%%%

%\section{Methodology LONG}
%\input{section_methods}

\section{Theory}

%\section{Methodology}

%\subsection{Theory}

Magnetostriction describes deformations of the material dimensions in response to external magnetic field. It has two contributions, the anisotropic magnetostriction originating from the dependence of magnetocrystalline anisotropy and the isotropic volume one based on the isotropic exchange interactions. 

Small deformations in solids are described by a strain tensor~\cite{landau1986theoryofelasticity}
\begin{equation}
    \varepsilon_{ij} = {\frac{1}{2}} \left( \frac{\partial u_{i}}{\partial r_{j}} + \frac{\partial u_{j}}{\partial r_{i}} \right) \, ,
\end{equation}
where  $\mathbf{u}(r)=\mathbf{r}^{\prime}-\mathbf{r}$ denotes a displacement vector.
The magnetostriction can be expressed by the magnetization given change of the length of an originally demagnetized material \cite{MAELAS_1_r21,Cullity_Graham,chikazumi2009physics}
\begin{equation}
   \frac{l-l_{0}}{l} \bigg{\vert}^{\bm{\alpha}}_{\bm{\beta}}  =  \sum_{i,j=x,y,z} \varepsilon^{{eq}}_{ij} {\beta}_{i} {\beta}_{j} \, ,
   \label{Eq.magstrictcoeff}
\end{equation}
where $l_{0}$ denotes the length along the $\bm{\beta}$ direction of a demagnetized sample and $l$ is the length in the same direction when the sample is magnetized along the $\bm{\alpha}$ direction.
The equilibrium strain tensor $\varepsilon^{{eq}}_{ij}$  comes from a minimization of the sum of the elastic energy $E_{el}$ and  magnetoelastic energy $E_{me}$ with the strain \cite{engdahl2000handbook,chikazumi2009physics,CLARK1980531}
\begin{equation}
    \frac{\partial  ( E_{el} + E_{me})}{\partial \varepsilon^{{eq}}_{ij}} = 0  \, .
\end{equation}

Performing the energy minimization %of the sum of elastic and magnetoelastic energy 
with respect to the strain, the magnetostriction given by magnetizing a sample, {\it e.g.} by an external field, can be obtained. The relative length change at the equilibrium strain in the tetragonal (I) system follows~\cite{MAELAS_1_r21}
\begin{align}
    & \left.\frac{\Delta l }{l_{0}} \right\vert^{\bm{\alpha}}_{\bm{\beta}}  = \lambda^{\alpha 1,0} (\beta_{x}^{2} + \beta_{y}^{2}) 
    + \lambda^{\alpha 2,0} \beta_{z}^{2} \label{Eq.rel_l_change} \\  &  +  \lambda^{\alpha 1,2} (\alpha_{z}^2-\frac{1}{3}) (\beta_{x}^{2} + \beta_{y}^{2}) + \lambda^{\alpha 2,2} (\alpha_{z}^2-\frac{1}{3}) \beta_{z}^{2}
    %\nonumber \\
    \nonumber  \\
    &  +\frac{1}{2}  \lambda^{\gamma,2} (\alpha_{z}^2-\alpha_{y}^{2}) (\beta_{x}^{2}  - \beta_{y}^{2}) + 2\lambda^{\delta,2} \alpha_{x}\alpha_{y}\beta_{x}\beta_{y} \nonumber \\
    \nonumber & + 2\lambda^{\varepsilon,2} (\alpha_{x}\alpha_{z}\beta_{x}\beta_{z} + \alpha_{y}\alpha_{z}\beta_{y}\beta_{z}) \nonumber \, ,
\end{align}

%%%%%%%%%%%%%%%%%%%
where the  magnetostrictive coefficients {$\lambda=\lambda(b_{i},C_{ij})$}  (Eqs.~\ref{Eq.L_a10}-\ref{Eq.L_e2}) are functions of the elastic coefficients $C_{ij}$ (Eq.~\ref{Eq.ElasticSI}) and magnetoelastic constants $b_{i}$ (Eq.~\ref{Eq.magel_const_SI})~\cite{maelas3_nieves_r23}. The first two $\lambda$ coefficients describe the  volume magnetostriction, whereas the rest ones are related to the magnetization direction dependent anisotropic magnetostriction.

{This theoretical description  was {originally} derived for a FM phase, however is it valid also for the assumed collinear AFM phases. The AFM structure can be split  into two antiparallel FM sublattices $\mu=\{A,B\}$, whose magnetoelastic energy satisfy \cite{CallenPRB1963}
\begin{equation}
    E_{me}^{\mathrm{total}} = E_{me}^{A} (\alpha_{A}) + E_{me}^{B} (\alpha_{B}); \quad \alpha_{A}=-\alpha_{B} \, ,
\end{equation}

\begin{align}
   &\frac{1}{V_{0}} E_{me}^{\mu} =b_{11}^{\mu}(\varepsilon_{\mathrm{xx}} + \varepsilon_{\mathrm{yy}})  + b_{12}^{\mu}\varepsilon_{\mathrm{zz}}  \\ & + b_{21}^{\mu} (\alpha_{z,\mu}^{2} - \frac{1}{3})(\varepsilon_{\mathrm{xx}} + \varepsilon_{\mathrm{yy}} ) +  b_{22}^{\mu} (\alpha_{z,\mu}^{2} - \frac{1}{3})\varepsilon_{\mathrm{zz}} \nonumber \\ & +  \frac{1}{2} b_{3}^{\mu} (\alpha_{x,\mu}^{2} - \alpha_{y,\mu}^{2}) (\varepsilon_{\mathrm{xx}} - \varepsilon_{\mathrm{yy}}) + 2 b_{3}^{\prime \mu} \alpha_{x,\mu} \alpha_{y,\mu} \varepsilon_{\mathrm{xy}}  \nonumber  \\ &+ 2 b_{4}^{\mu} ( \alpha_{x,\mu} \alpha_{z,\mu} \varepsilon_{\mathrm{xz}} + \alpha_{y,\mu} \alpha_{z,\mu} \varepsilon_{\mathrm{yz}}) \nonumber \, .
\end{align}
Thanks to the opposite spin direction ($\alpha_{A}=-\alpha_{B}$), the  magnetoelastic energies of the sublattice A and B are same 
\begin{equation}
    E_{me}^{A} (\alpha_{A}) = E_{me}^{A} (-\alpha_{A}) = E_{me}^{B} (\alpha_{A}) \, ,
\end{equation}
which yields for the magnetoelastic constants $b_{i}$
\begin{equation}
    b_{i} =2b_{i}^{A}=2b_{i}^{B}
\end{equation}
assuming scaling to the total volume $V_{0}$. 
}

Considering a polycrystal, a mean fractional length change between an initial state and a saturated final state can be evaluated by averaging through the magnetization directions. It yields relations  as follows \begin{equation}
   \left\langle \frac{l-l_{0}}{l} \bigg{\vert}^{\bm{\alpha}}_{\bm{\beta}}  \right\rangle =  \xi + \eta(\bm{\alpha}\cdot\bm{\beta})^{2} \, ,
   \label{Eq.polycrys_rel_l_chage}
\end{equation}
where the form of  $\xi$, $\eta$ parameters depends on the initial demagnetized state~\cite{MAELAS_2_r22}. 
Assuming an initial state with domains aligned along the easy direction, the $\xi$ and $\eta$ parameters are given as follows 
\begin{align}
\xi = &  \frac{4}{15} \lambda^{\alpha1,2} + \frac{1}{15} \lambda^{\alpha2,2} - \frac{2}{15} \lambda^{\epsilon,2} -\frac{1}{15} \lambda^{\gamma,2} -\frac{1}{15} \lambda^{\delta,2} \nonumber \\  & -\frac{1}{l3} (2 \lambda^{\alpha1,2}  +  \lambda^{\alpha2,2})\, \mathrm{cos}^{2} \Omega \, , \\
\eta = &-\frac{2}{15} \lambda^{\alpha1,2} + \frac{2}{15} \lambda^{\alpha2,2} + \frac{2}{5} \lambda^{\epsilon,2} + \frac{1}{5} \lambda^{\gamma,2}  + \frac{1}{5} \lambda^{\delta,2} \, .
\label{Eq.xi_domains}
\end{align}
where $\Omega=0$ for easy axis and $\Omega=\frac{\pi}{2}$ for easy plane system, respectively.  %T

\FloatBarrier

\section{Experimental}
% experimental

{To experimentally measure the magnetostriction,}
a polycrystalline sample of MnPt was prepared using an arc melting process with the MAM-1 system (Edmund Bühler GmbH). Stoichiometric amounts of high-purity manganese (99.9\%) and platinum (99.99\%) were melted under a titanium-gettered argon atmosphere. To ensure homogeneity, the sample was flipped and remelted multiple times. The final sample weighed approximately 1~g, with a mass loss of less than 0.5\%. No further heat treatment was applied.
X-ray diffraction (XRD) measurements were conducted at room temperature on a sample that had been hand-ground. These measurements were performed using a PANalytical X’pert Pro diffractometer, employing CuK$\alpha$ radiation produced at 40 kV and 30 mA ($\lambda = 1.5406$ Å) in a Bragg-Brentano geometry. The resulting XRD pattern was analyzed using FullProf software \cite{RODRIGUEZCARVAJAL199355} (Fig.~\ref{fig:xrd-pattern}). 
High-resolution magnetostriction measurements, tracking length changes as a function of magnetic field (magnetostriction) at 2~K, were performed using a miniature capacitance dilatometer~\cite{Rotter_r98_magnetometer}. %
The dilatometer was connected to an AH2500A capacitance bridge, integrated into a Physical Property Measurement System (PPMS) from Quantum Design.

\section{Calculation details}
% calculations

\begin{figure*}[t]
\centering
\includegraphics[width=0.65\textwidth]{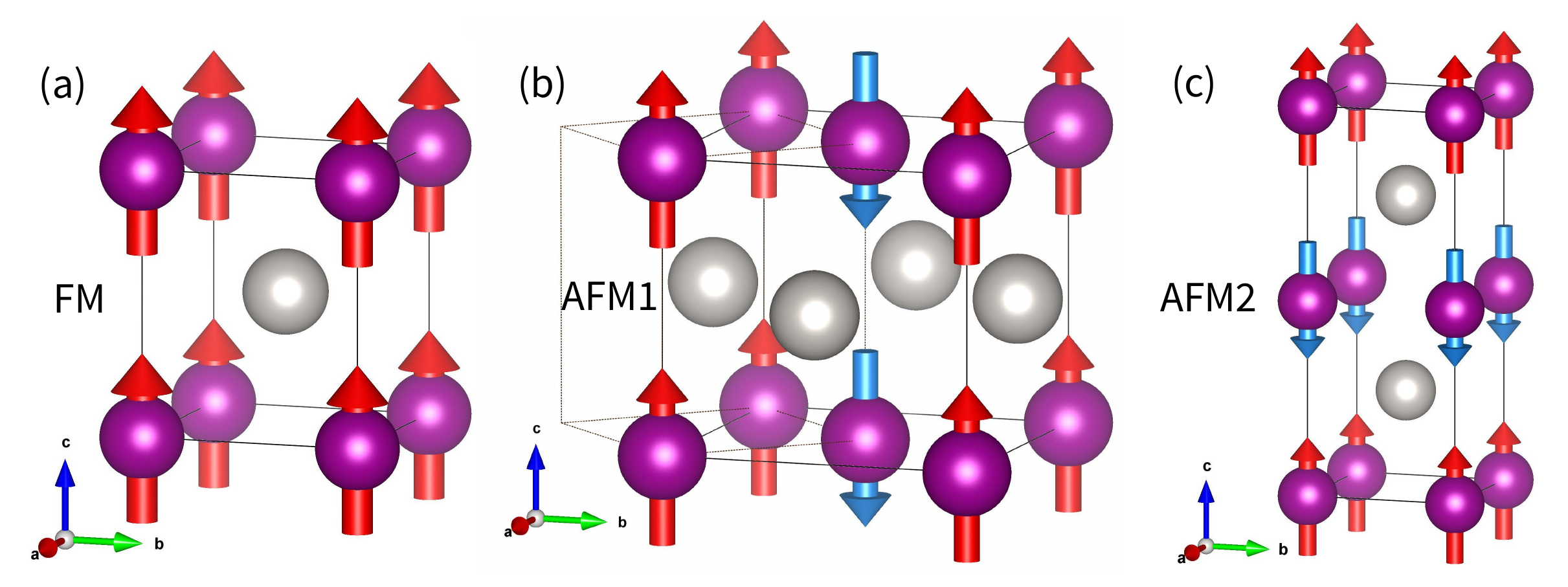}
\caption{\label{fig:Mag_structures} MnPt magnetic structures. (a) FM, (b) AFM1, (c) AFM2. Dashed lines in the AFM1 structure denote the primitive cell, similar to the FM one. (plotted in VESTA 3~\cite{VESTA})}
\label{Fig.:magstructure}
\end{figure*}

Electronic structure calculations were performed within the plane-wave based Vienna ab-initio simulation package (VASP)~{\cite{Kresse_VASP_r96,Kresse_VASP_r99}} employing the projector-augmented-wave (PAW) method with PAW pseudo-potentials. Calculations consider the generalized gradient approximation (GGA) of Perdew-Burke-Ernzerhof (PBE)~\cite{PBE_r96}. Primarily, the non-collinear magnetic calculations with the spin-orbit coupling were considered. The energy cut-off  for the plane waves of 450 eV was used. In general, an automatic generation of a $k$-mesh scheme with $R_{k}$~=~70 was considered. %  
The relaxation was performed in the Methfessel-Paxton scheme of the order~1 (Brillouin zone (BZ) integration technique) with a smearing 0.01~eV. Due to computational demands, the same smearing was used when $R_{k}$ exceeded 100. Elsewhere, 
the tetrahedron method with Blöchl corrections was employed for the BZ integration. The energy convergence in structure relaxation calculations includes the convergence of the self-consistent loop better than EDIFF~=~10$^{-7}$~eV and energy difference between relaxation steps smaller than EDIFF~=~10$^{-6}$~eV. Regarding calculations of the magnetoelastic properties and MAE, a tight convergence was demanded EDIFF~=~10$^{-9}$~eV.
To obtain a good linear fit of magnetoelastic constants $b_{i}$, $R_{k}$~=~120 resp. $R_{k}$~=~110 was used for the calculations of elastic and magnetoelastic properties in the case of AFM1, resp. AFM2 magnetic structures (Fig.~\ref{fig:Mag_structures}b,c). The elastic and magnetoelastic parameters were evaluated within AELAS~\cite{AELAS_r17} resp. MEALAS {(with -mode 2, strain-energy method)}~\cite{MAELAS_1_r21,MAELAS_2_r22} packages using a finite displacement approach. The packages were used to generate distorted structures and the result analysis.

The applied strain in the magnetoelastic calculations differs between magnetic phases depending on the linearity of the obtained energy dependencies. In the case of the FM structure (Fig.~\ref{fig:Mag_structures}a), the maximum applied strain $\varepsilon$ was 0.0075. The AFM1 system exhibits linear behavior in a smaller region, and the maximal strain was reduced to 0.0050. Whereas for the AFM2 structure, the maximum strain 0.0100 can be applied except for the calculation of the $b_{2}$ parameter, requiring a reduction to 0.0075. The R-factor of the linear fitting was better than 0.98 except for certain cases of low $b_{i}$ values given by related extreme demands on precision. %

MAE was estimated at a fine k-mesh (FM: $R_{k}$~=~70, AFM1: $R_{k}$~=~90 and AFM2: $R_{k}$~=~95) with tight energy convergence EDIFF~=~$10^{-8}$~eV. High-accuracy calculations were needed to obtain smooth energy curves, particularly for antiferromagnetic structures.

The exchange interaction parameters to atomistic spin simulations were calculated within the Relativistic Spin Polarized toolkit (RSPt) package based on the Full-Potential Linear Muffin-Tin Orbital (FP-LMTO) method~\cite{Wills2010_lmto_rspt,Kvashnin_prb91_r15}.  The calculations were performed on 20$\times$20$\times$20 $k$-mesh in the fully relativistic scheme, considering the xc-potential of PBE 96~\cite{PBE_r96} and Perdew Wang 1992~\cite{PW92_prb45_r92}. Relaxed VASP crystal structures were used in the calculations. The energy convergence was better than 10$^{-10}$~ Ry and the orbitals $s$, $p$, $d$ were considered in the evaluation of exchange interactions.
The isotropic magnetic exchange interactions $J_{ij}$ were evaluated by the fully relativistic version of the LKAG method based on the magnetic force theorem~\cite{Szilva_r23_RevModPhys}. The interaction Hamiltonian reads 
\begin{equation}
\hat{H} = - \sum_{i\neq j} \sum_{\{\alpha,\beta\} =\{x,y,z\} }  e_{i}^{\alpha}  \hat{J}_{ij}^{\alpha\beta} e_{j}^{\beta} \,   ,
\end{equation}
where  $\hat{J}_{ij}^{\alpha\beta}$ is a (3$\times$3) interaction coupling tensor and $e_{i}^{\alpha}$ denotes spin components of unitary spin vector  at the site $i$. However,  here, only  the AFM1 phase related isotropic Heisenberg interactions $J_{ij}$ (Fig.~\ref{fig:Jij_volume-dep}), given as follows~\cite{Szilva_r23_RevModPhys}
\begin{align}
    \hat{H} = - \sum_{i\neq j}   {J}_{ij}   \mathbf{e}_{i}\cdot\mathbf{e}_{j} \, , \\
    J_{ij}=\frac{1}{3}\mathrm{Tr}\hat{J}_{ij}^{\alpha\beta} \, ,
\end{align}
were considered in the atomistic spin simulations together with the calculated anisotropic constants $K_{1}$ and $K_{2}$. They  were employed to simulate   magnetization directions of magnetic sublattices with respect to the applied external magnetic field.
Atomistic spin simulations were  performed  within the Uppsala Atomistic Spin Dynamics (UppASD) program~\cite{Skubic_jpcm_r08_uppasd}, namely, a magnetization averaged over time and particular magnetic sublattices of the AFM1 system was studied. 
To describe a polycrystalline system, a few distinct magnetic field orientations with respect to the crystal structure were considered, i.e., the magnetic field along the $x$, $z$, $xz$, and $xy$ Cartesian directions, where the AFM1  crystal structure $a$ axis is oriented along the Cartesian $x$ direction and the $c$ axis coincides with the $z$ direction. A 20x20x20 supercell with Mn atoms,  bearing the magnetic moment as in the AFM1 phase, was considered. The simulations were performed at $T=2$~K according to the experiment, while the time and Mn atoms averaged magnetization were studied.

%%%%%%%%%%%%%%%%%%%%%%%%%%%%%%%%%%%%%%%%%%%%%%%%%%%%%%%%%%%%%%%%%%%%%%%%%%%%%%%%%%%
%  Results 
%%%%%%%%%%%%%%%%%%%%%%%%%%%%%%%%%%%%%%%%%%%%%%%%%%%%%%%%%%%%%%%%%%%%%%%%%%%%%%%%%%%

\section{Results}
%  Results 

%\subsection{Magnetic structures}
%

%fig 1

\subsection{Theoretical magnetoelastic behavior}

{Prior to the analysis of the experimental  magnetostriction results, theoretical magnetoelastic properties are estimated.}
Despite the experimentally reported antiferromagnetic ground state of the studied MnPt system~\cite{pal_jap_r68}, {in our calculations},  we also considered other magnetic ordering~\cite{QURATULAIN_jmmm_r18,Lu_prb_r10}  to inspect the 
the magnetoelastic behavior {and analyze the experimental behavior}. 
Namely, we considered three collinear structures with magnetic moments pointing along the $c$-axis:   ferromagnetic (FM) and two antiferromagnetic (AFM1, AFM2) structures (Fig.~\ref{Fig.:magstructure}). % 
Corresponding to the literature~\cite{pal_jap_r68,Lu_prb_r10,Kang_prb_r23,QURATULAIN_jmmm_r18}, the AFM1 state, with AFM ordered nearest-neighbor spins in the basal $ab$-plane, was found as the ground state (Table~\ref{Tab:SI1_en_str}).
However, it should be noted that the FM phase was observed in quenched powders and sputtered films of disordered MnPt~\cite{Severin_jap_r78}. {The additional AFM 2 phase serves  for comparison with another AFM phase.}
The obtained energy differences are in agreement with the published ones~\cite{QURATULAIN_jmmm_r18,Lu_prb_r10} as well as the relaxed structural parameters~\cite{WANG_jmmm_r13,Lu_prb_r10,Kang_prb_r23,Ravindran_PRB_01}  (Table~\ref{Tab:SI1_en_str}). %

\begin{table}[t]
\centering

\caption{\label{Tab:multi_tab}%
MnPt magnetoelastic constants $b$, magnetostrictive coefficients $\lambda$, and magnetostriction parameters for a polycrystalline sample $\xi$, $\eta$  . Axes are oriented according to the FM primitive cell, i.e., the AFM1 structure is rotated about 45 degrees along the c-axis .}
\begin{tabular}{cccccc}
\hline
\hline
$b$ (MPa) &  FM & AFM1 &  AFM2\\
\hline
$b_{\mathrm{21}}$ & 135  & 12   & -62  \\
$b_{\mathrm{22}}$ & -111 & -19  & 23  \\
$b_{\mathrm{3}}$ &  -40 & -37  & 118  \\
$b_{\mathrm{4}}$ &  -34& 1  & 26  \\
$b_{\mathrm{3}}^{\prime}$ & 86  & 47  & 75  \\

\hline
\hline

$\lambda$ (10$^{-6}$) &  FM & AFM1 & AFM2\\

\hline
$\lambda^{\mathrm{\alpha 1,2}}$ & -1592  & -88   & 757   \\
$\lambda^{\mathrm{\alpha 2,2}}$ &  2655 & 155  & -1168   \\
$\lambda^{\mathrm{\gamma, 2}}$ & 240   & 156   & -754   \\
$\lambda^{\mathrm{\varepsilon, 2}}$ & 170  &  -6  & -151   \\
$\lambda^{\mathrm{\delta, 2 }}$ & -620  & -340   & -775   \\
\hline
\hline

 &  FM & AFM1  & AFM2 \\
\hline
$\xi$ (10$^{-6}$) & -68 & 7  &  130   \\
$\eta$ (10$^{-6}$) & 558& -7  &  -623 \\
\hline
%\hline
\end{tabular}
\end{table}

Having calculated  the elastic constants (Table~\ref{Tab:SI3_elast}), the  {magnetoelastic ($b_{i}$) and magnetostrictive coefficients ($\lambda_{i}$)  can be discussed , respectively.}
The magnetoelastatic constants arise from the change of the magnetocrystalline anisotropy energy (MAE)  with respect to the applied strain and the defined spin directions (Fig.~\ref{fig:magel_chard}).
The type of magnetic order substantially  modifies the MAE (Fig.~\ref{fig:MAE}).  Thereby,  the magnitudes of the $b_{i}$  parameters, including their signs, depend strongly on the magnetic structure as demonstrated by the estimated values of the magnetoelastic constants $b_{i}$ (Table~\ref{Tab:multi_tab}). For clarity, the values of $b_{i}$ constants in  are referred to the axis of the FM structure, similar to the  case of the elastic constants. 
The $b_{i}$ constants differ significantly, except {for} the  $b_{3}^{\prime}$ one.

%%%%%%%%%%%%%%%%%%%%%%%%%%%%%%%%%%%%%%%%%%%%%%%%%%%%%%%%%%%

More important from an experimental point of view are the magnetostrictive coefficients $\lambda^{i}$ (Table~\ref{Tab:multi_tab}) that describe the change in length with respect to the sample magnetization (Eq.~\ref{Eq.magstrictcoeff}). Magnetic structure-dependent differences in magnetoelastic constants $b_{i}$ give rise to substantial differences in magnetostrictive behavior. The most striking {ones} are in the magnitudes of the $\lambda^{{\alpha} 1,2}$ and $\lambda^{{\alpha} 2,2}$ constants (Eqs.~\ref{Eq.L_a12},\ref{Eq.L_a22} ) coming from the $b_{21}$ and $b_{22}$ magnetoelastic coefficients. The constant $\lambda^{{\alpha} 1,2}$ describes an enlargement of the $ab$-basis  area for magnetization applied along the $c$-axis, while $\lambda^{{\alpha} 1,2}$ is related to the elongation of the $c$-parameter. 
Concerning the FM state, a huge  $ab$-basis area squeezing {was revealed}, being compensated by enormous $c$-axis elongation. An opposite behavior, about half-magnitude weaker, was found for the AFM2 phase. The opposite signs come from the magnetoelastic coefficients  {and they} are also responsible for modest values related to the AFM1 phase, since the elastic constants are comparable between magnetic {phases}.
Found magnitudes of FM and AFM2 coefficients  $\lambda^{{\alpha} 1,2}$ and $\lambda^{{\alpha} 2,2}$   are outstanding compared to other compounds~\cite{Nieves_sss_r25_seconorder}. 

The $\lambda^{{\gamma}, 2}$ coefficient (Eq.~\ref{Eq.L_g2} arising from $b_{3}$ constant denotes the change in length along the $a$ resp. $b$ axis with difference in the magnetization direction $(\alpha_{x}^{2}-\alpha_{y}^{2})$ resp. $(\alpha_{y}^{2}-\alpha_{x}^{2})$. Hence, it does not contribute when $\alpha_{x}^{2}=\alpha_{y}^{2}$. Both the FM and AFM1 states in the basal plane prefer the elongation along the basal magnetization. The behavior of the AFM2 is opposite and way stronger. 
The shear in the $ab$ basis given by the in-plane components of the magnetization axis is attributed to the $\lambda^{{\delta}, 2}$ coefficient (Eq.~\ref{Eq.L_d2}).  It is substantial irrespective of the magnetic order, bearing similar values {FM and AFM2 phases} as {their} $b_{3}^{\prime}$ hardly vary.
On the other hand, the shear perpendicular to the $ab$ basis related to $\lambda^{{\varepsilon}, 2}$ (Eq.~\ref{Eq.L_e2}) is much weaker. Particularly for the AFM1 phase, the MAE is almost unchanged with deformation ($b_{4}$).
%%%%%%%%%%%%%%%%%%%%%%%%%%%%%%%%%%%%%%%%%%%%%%%%%%%%%%%%%%%%

\subsection{Experimental magnetostriction analysis}
 
\begin{figure*}[t]
    \centering
    \includegraphics[width=0.85\textwidth]{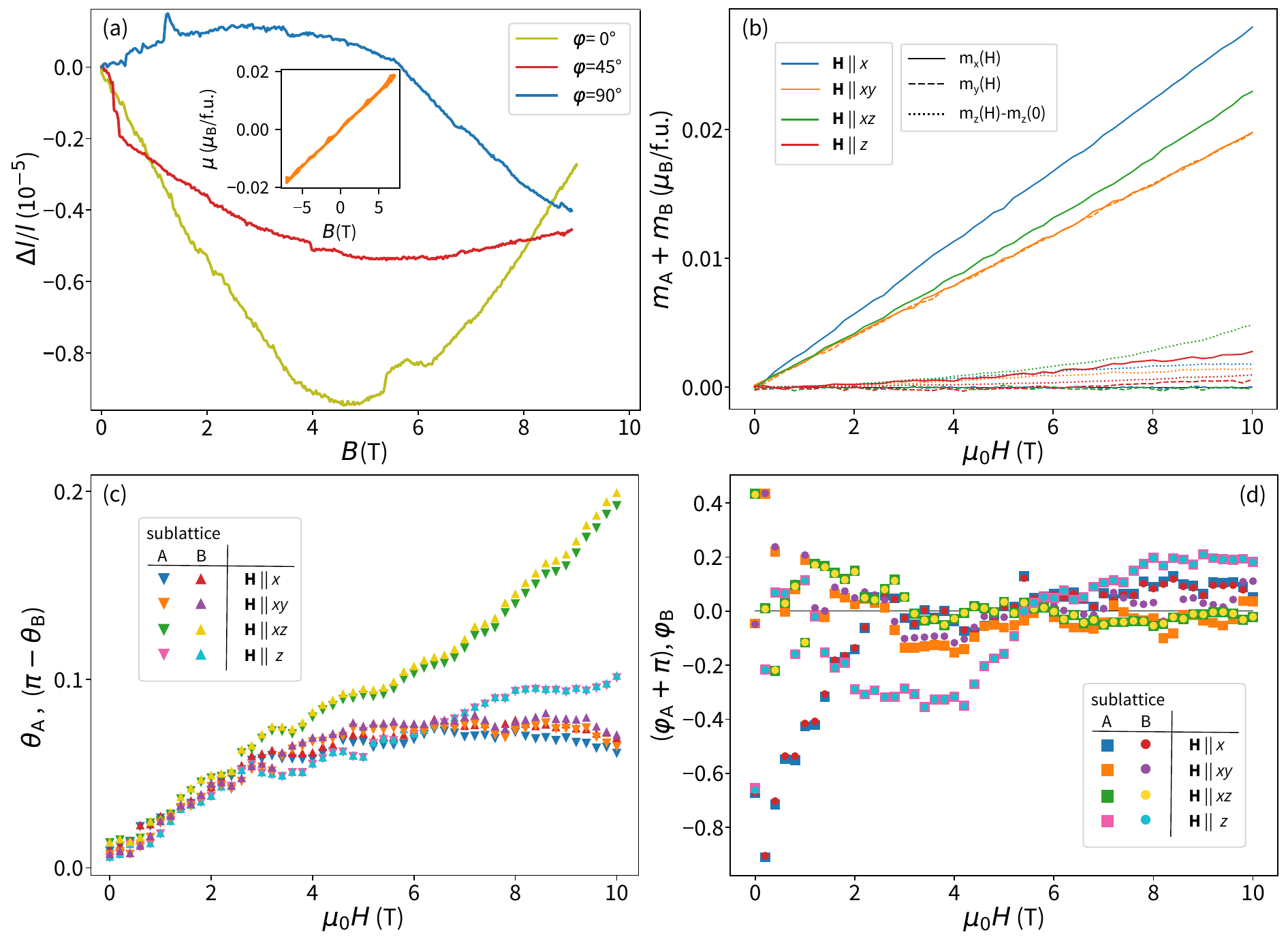}
   \caption{Experimental MnPt magnetostriction and simulated sublattice magnetization directions.  (a) Magnetostriction  measured (olive) parallel, (blue) perpendicular, and (red) at 45 degrees to the applied external magnetic field. The inset depicts the magnetization curve of the sample. (b-d)  magnetization direction of magnetic sublattices A and B simulated in the AFM1 system depending on the external field strength. (b) Cartesian components of the total magnetization. (c,d) Magnetization directions related to the A,B sublattices in spherical coordinates.  Magnetization is averaged over the atoms in the particular supercell  and the time.  Four different relative field orientations were applied. The $a$ axis of the AFM1 crystal cell is oriented along the Cartesian $x$ direction, and the $c$ axis along the $z$ direction.   Calculations were performed at $T=2$~K same as the experiment.
    }
    \label{fig:simul_magnetiz}
\end{figure*}

{The obtained theoretical results are used to explain the measured behavior of a polycrystalline sample.}
The measured dilatometry (Fig.~\ref{fig:simul_magnetiz}) was performed with different mutual orientations of the length change direction and the applied external magnetic field.
Regarding the measured data for the parallel field orientation ($\varphi=0$ deg),   the sample length shrinks up to the field of $\mu_{0}H\sim5$~T. However, further on, the sample starts to elongate, as the opposite slope of the measured dependence occurs.
Applying a field perpendicular to the measured direction, a slight sample elongation was detected for weak fields. Increasing the field strength, the slope changes around a field magnitude similar to the case of the parallel field direction, and the measured sample direction starts to be reduced. 

%fig 2

The observed behavior is related to magnetizing a sample of the AFM nature, where the behavior is more complex than in the FM system due to the AFM ordering.  Performing the measurements at 2~K, one can assume that the AFM domains are randomly oriented along the easy axis directions, as the measurements are well below the N\'{e}el temperature.
Applying the external field, in a textbook model, the AFM magnetic structure is modified according to the orientation of the easy axis in the magnetic domain with respect to the external magnetic field direction. For simplicity, let us consider only two limit orientations, i.e., external field parallel to the easy axis and the perpendicular orientation of the easy axis and the field.
Regarding the perpendicular orientation, the external field introduces canting of the AFM moments into the direction of the magnetic field. Whereas for the parallel orientation, a spin-flop meta-magnetic transition at a certain field magnitude should appear as the system possesses weak anisotropy. In general, the applied external field smears out the AFM character of the system. On the other hand, in the FM system, the magnetic structure is kept despite the tilting of the magnetic moments.

\begin{figure*}[t]
\includegraphics[width=\textwidth]{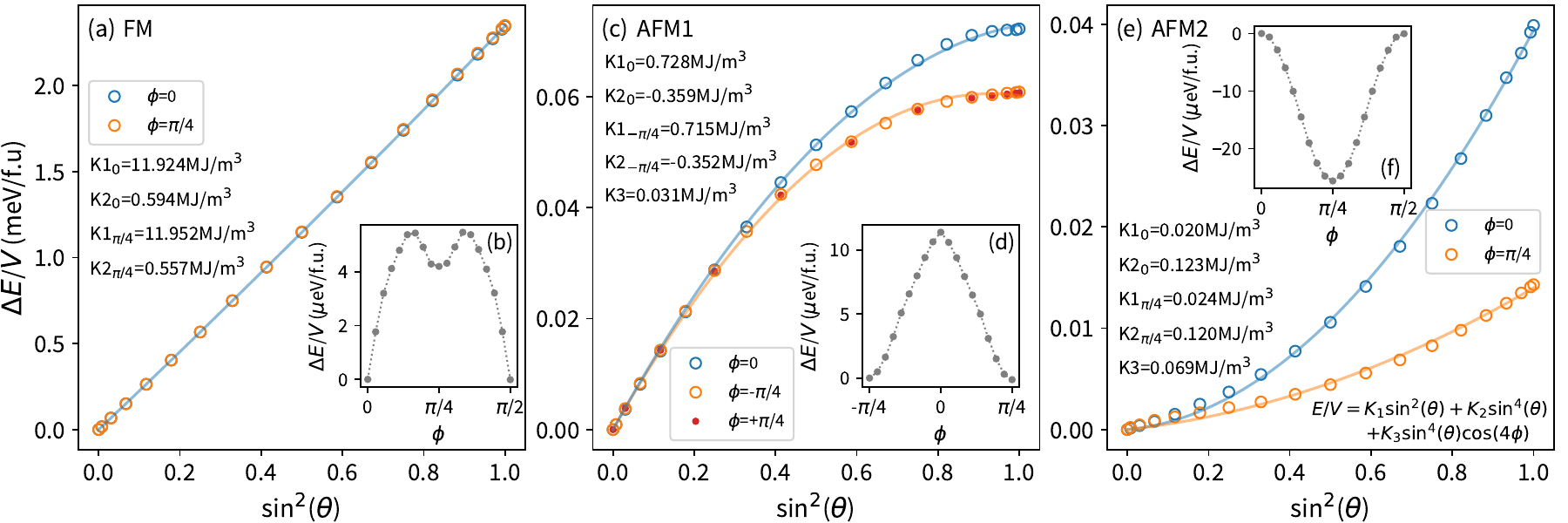}
\caption{\label{fig:MAE} Magneto-crystalline anisotropy. (a,b) FM, (c,d) AFM1, (e,f) AFM2 magnetic phases. The insets (b,d,f) denote a change of the MAE in the ab-plane. Regarding the AFM1 magnetic structure, the axis orientation according to the  FM structures is considered. 
}
\end{figure*}

\begin{figure*}
\includegraphics[width=0.99\textwidth]{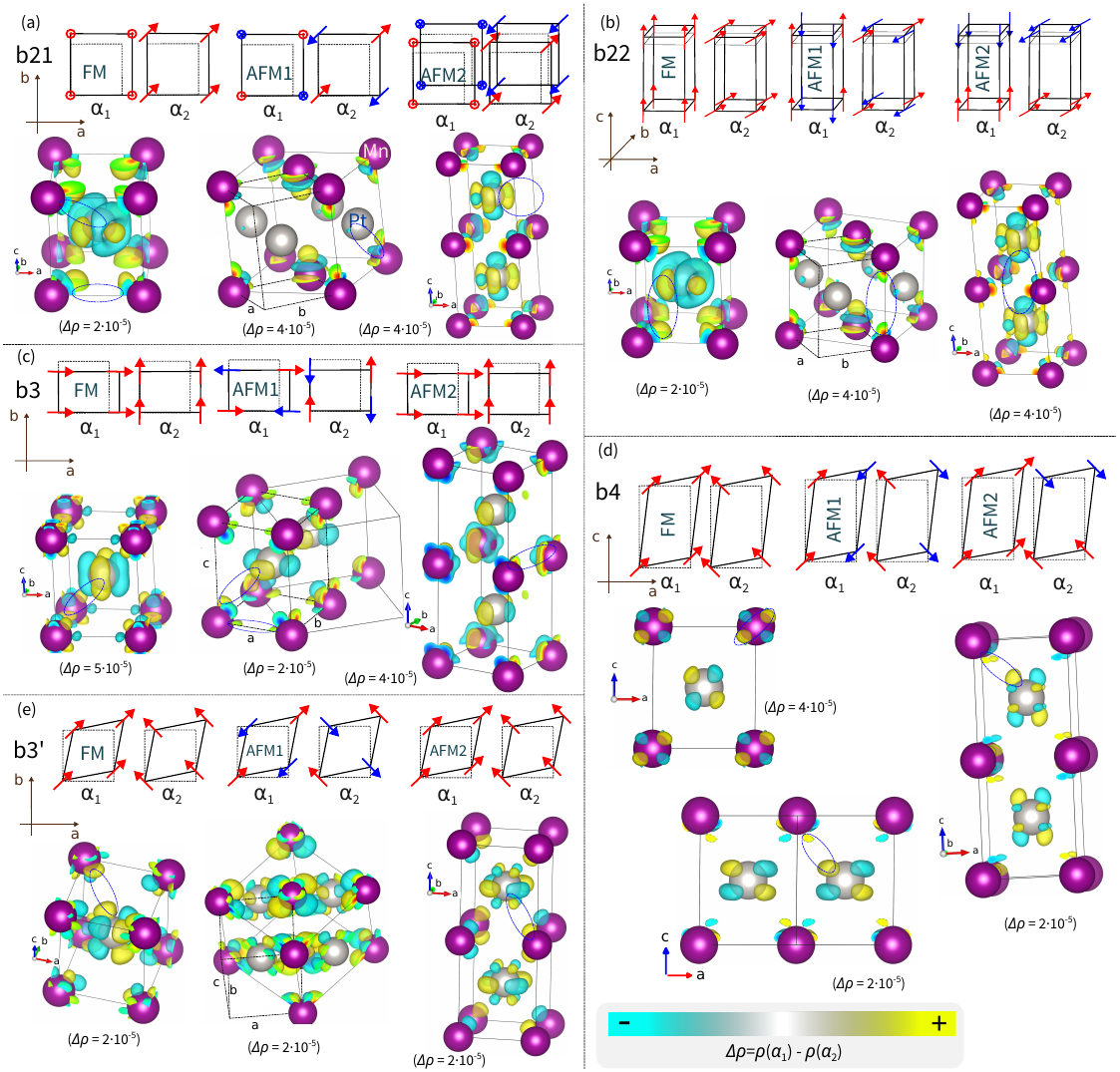}
\caption{\label{fig:magel_chard} Lattice deformations and spin directions related to calculations of magnetoelastic coefficients with related charge density differences. For each determined magnetoelastic coefficient $b_{i}$, lattice deformations applied to the FM, AFM1, and AFM2 magnetic structures are shown with depicted spin orientations $\alpha_{1}$ and $\alpha_{2}$.   Charge density difference between the system with magnetization along $\alpha_{1}$ and $\alpha_{2}$ directions from self-consistent calculations is shown for each magnetic phase and type of deformation with the applied strain $\varepsilon$=0.005. {Fine k-mesh was used (FM: $R_{k}$~=~70, AFM1: $R_{k}$~=~90 and AFM2: $R_{k}$~=~95).} Yellow color denotes an excess of the charge density difference related to the $\alpha_{1}$ magnetization direction, whereas cyan one to the $\alpha_{2}$ direction.  Deformations with respect to the FM axes are considered as in the Table~\ref{Tab:multi_tab}.  Below each charge density plot, the magnitude of the plotted $\Delta\rho$ isosurface is stated. Charge density plotted in VESTA 3~\cite{VESTA}. }
\end{figure*}

However, to obtain an actual response of the magnetic structure to the applied external field, atomistic spin simulations of the magnetization direction were performed.
%
%Spin simulations of the magnetization
They show that the canting of the magnetic moments is small (Fig.~\ref{fig:simul_magnetiz}c,d). The induced magnetization is negligible (Fig.~\ref{fig:simul_magnetiz}b), and the simulations correspond well to the experimental values (Fig.~\ref{fig:simul_magnetiz}a-inset). 
Analysis of the sublattice magnetization direction revealed that, particularly for low fields, the sublattice magnetizations point in nearly opposite directions (Fig.~\ref{fig:simul_magnetiz}c,d), as the magnetization directions are almost simultaneously tilted, resembling the behavior of a FM system. Moment canting is negligible, which seems to arise from a low magnetocrystalline anisotropy (Fig.~\ref{fig:MAE}b).

It enables one to characterize the AFM1 system  at low fields in a similar way to the FM one. Therefore, we calculated the parameters $\xi$ and $\eta$ (Table~\ref{Tab:multi_tab}) describing a polycrystalline system according to Eq.~\ref{Eq.polycrys_rel_l_chage} for both AFM1 and FM MnPt magnetic phases. 
Regarding a weak parallel external field, where the canting is negligible (Fig.~\ref{fig:simul_magnetiz}c,d), the negative measured magnetostriction (Fig.~\ref{fig:simul_magnetiz}a) corresponds to the sign of the calculated AFM1 $\eta$ parameter. Actually, one can ascribe it to the difference between the parallel and perpendicular field setups in the measurements.  Since saturation cannot be achieved as in the FM case, it is hard to compare the magnitude. The change of the slope and occurrence of the positive magnetostriction in the experimental measurements with increasing external field can be attributed to the canting of the magnetic moments as the AFM magnetic structure starts to be substantially modified (Fig.~\ref{fig:simul_magnetiz}c,d) according to the simulations between 4~T and 6~T, in agreement with the measurements. Assuming a sort of FM-like contribution, the behavior will agree with the calculated positive FM $\eta$ parameter being nearly five times stronger than the AFM1 one (Table~\ref{Tab:multi_tab}).   Concerning the $\xi$ parameter, one can focus on the measurements with the perpendicular field orientation (Eq.~\ref{Eq.polycrys_rel_l_chage}).  Following the aforementioned assumptions, the sign of the AFM1 resp. FM $\xi$ parameters (Table~\ref{Tab:multi_tab})  corresponds to the observed behavior under weak and high external magnetic field (Fig.~\ref{fig:simul_magnetiz}).

%\FloatBarrier

%fig 3

%%%%%%%%%%%%%%%%%%%%%%%%%%%%%%%%%%%%%%%%%%%%%%%%%%%%%%%%%%%

%fig 4

\subsection{Magnetic structure dependent mangetoelastic behavior}

The calculated magnetic structure dependence  of the $b_{i}$ constants, which represent differences in the slopes of the MAE  with respect to the applied strain~\cite{MAELAS_2_r22}, can be understood starting from differences in spin structure in relation to the magnetization axes' directions, or more carefully by induced charge density differences regarding the applied deformations (Fig.~\ref{fig:magel_chard}).

Begining from the $b_{21}$ constant, the area of the $ab$-base is changed (Fig.~\ref{fig:magel_chard}a). Enlarging the base area, the $c/a$ ratio gets closer to 1, and the MAE for the defined magnetization directions $\alpha_{1}$ and $\alpha_{2}$ should be smaller by simple consideration. It agrees with the positive  $b_{1}$ sign, except AFM2 case, as the difference $E_{\alpha_{1}} -  E_{\alpha_{2}}$ is less negative. Concerning AFM2 and the related  $b_{21}$ minus sign, simply based on the spin orientations, one can find that both in the FM and AFM1 case, the spins along the magnetization axes $\alpha_{1}$ resp. $\alpha_{2}$ have the same direction, whereas for AFM2 it changes. It is the opposite along the $\alpha_{1}$  direction, but the same along the $\alpha_{2}$ one. A better explanation can be provided by changes in the charge density induced by the change of the magnetization axis, which were obtained by self-consistent calculations (Fig.~\ref{fig:magel_chard}a). 

Regarding the AFM2 magnetic phase, significant positive charge difference $\Delta\rho=\rho(\alpha_{1})-\rho(\alpha_{2})$ (Fig.~\ref{fig:magel_chard}a - yellow color) appears along $a$ and $b$ directions between Mn resp. particularly in between Pt atoms. Enlarging the $a$ parameter, the extra energy contribution related to the charge difference
for the $\alpha_{1}$ magnetization direction 
is reduced, which lowers the  $E_{\alpha_{1}}$ with the respect to $E_{\alpha_{2}}$. 
The nearest Mn-Pt distance is shorter than the $a$ parameter, so the effect might be more important. However, the change of the charge density has no dominating character in this direction.
On the other hand, quite the opposite character of the charge density differences for the FM magnetic phase was calculated. It agrees with the opposite sign of the $b_{21}$. 
Finally, the AFM1 charge difference does not offer an easily visible preference either for $\alpha_{1}$ resp. $\alpha_{2}$ magnetization direction, which might explain the small $b_{21}$ constant. Its positive sign likely comes from a negative charge density difference along the Mn-Pt direction. 
The parts of charge density contributing to the parameters $b_{i}$ are denoted in Fig.~\ref{fig:magel_chard}.

Qualitatively, the high FM $b_{21}$ value corresponds to a strong magneto-crystalline anisotropy in the FM phase compared to AFM ones (Fig.~\ref{fig:MAE}). The FM anisotropic constant $K_{1}$ is one or two orders of magnitude larger than those of AFM1 or AFM2, respectively.  Further, also $K_{2}$ constants differ significantly across the magnetic phases, being largest for the FM magnetic phase. 
The $K_{1}$ dominates for the FM state. However, {for AFM1 phase, the $K_1$ is comparable to $K_{2}$} and even more remarkably, the $K_{2}$ constant dominates the AFM2 state, making its behavior quite different. In addition, the $K_{3}$ constant is substantial in AFM states, while it is negligible for the FM state.  
The $b_{22}$ constant is related to the opposite behavior of the $b_{21}$ as it comes from elongation of the $c$-axis (Fig.~\ref{fig:magel_chard}b)  followed by an increase of the MAE  for the FM magnetic phase (Fig.~\ref{fig:MAE}). Indeed, we observed  opposite behavior both in the signs and magnitudes following the difference in charge densities (Fig.~\ref{fig:magel_chard}b).

%fig 4
\begin{figure*}[t]
\includegraphics[width=\textwidth]{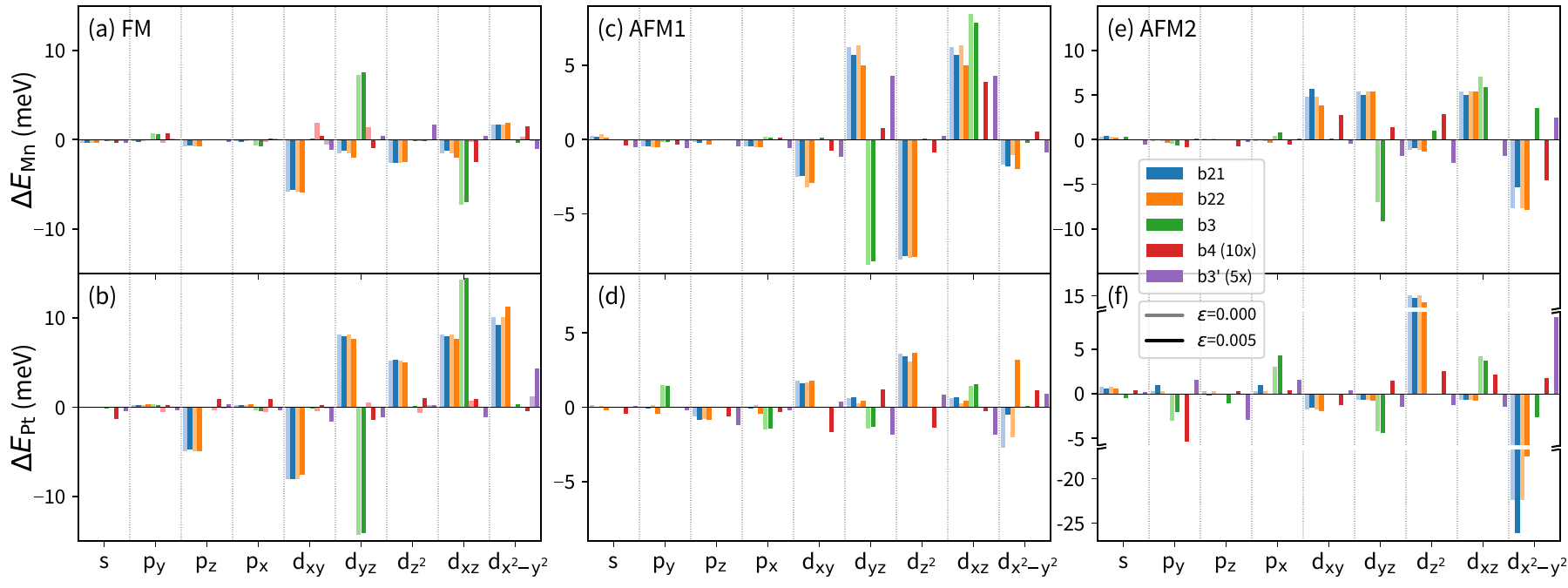}
\caption{\label{fig:EDiff_MAE} Atomic orbital resolved energy contributions to MAE. (a,b) FM, (c,d) AFM1, (e,f) AFM2. The energy difference  $\Delta E = E_{\alpha_{2}}-E_{\alpha_{1}}$ is related to magnetization axes as shown in  Fig.~\ref{fig:magel_chard}.}
\end{figure*}

Analyzing the orbital origin of the MAE, the shape of the charge differences and its sign correspond to the orbital resolved MAE contributions (Fig.~\ref{fig:EDiff_MAE}) obtained by summation of the orbital-resolved band energies across the Brillouin zone (Eq.~\ref{Eq.En_orbit_projection}).
The orbital-resolved  energy of the ion $i$  with magnetization axes $\alpha$ reads
\begin{equation}
    E_{lm}^{i}(\alpha) = \sum_{n,\mathbf{k}} E_{n\mathbf{k}} \vert \langle Y_{lm}^{i} \vert \phi_{n\mathbf{k}} \rangle \vert^{2} c_{n\mathbf{k}} \, ,
    \label{Eq.En_orbit_projection}
\end{equation}
where the $E_{n\mathbf{k}}$ is the band energy of the band $n$ at k-point~$\mathbf{k}$ with occupancy $c_{n\mathbf{k}}$. Finally, $Y_{lm}^{i}$ denotes spherical harmonics centered at the ion $i$.

Regarding the FM phase, there is a strong anisotropy coming from  Pt d-orbitals. Nevertheless, their contributions partly counteract. Besides, there are smaller contributions from Mn. Changing the magnetic structure, the Mn contribution dominates for the AFM1 phase corresponding to the charge density differences (Fig.~\ref{fig:magel_chard}b). The 
prominent $d_{xz}$ and $d_{yz}$ orbitals (Fig.~\ref{fig:EDiff_MAE}c,d) are related to negative $\rho$ which would explain positive $b_{21}$. A similar explanation applies to $b_{22}$ and $d_{z^{2}}$  (Figs.~\ref{fig:magel_chard}b and~\ref{fig:EDiff_MAE}c,d). 
Finally, Pt $d_{z^{2}}$ and $d_{x^{2}-y^{2}}$ contributions (Fig.~\ref{fig:EDiff_MAE}e,f) seem to be the most important for the structure of AFM2, corresponding to $\Delta \rho$ (Fig.~\ref{fig:magel_chard}a,b).   Both the orbital-resolved energy and charge density differences indicate that different orbitals are substantial depending on the magnetic structure, which seems to lead to different $b_{i}$ values.
One has to point out that the performed projections to the spherical harmonics (Eq.~\ref{Eq.En_orbit_projection}) are not complete, {and furthermore, they miss the interstitial charge. Thereby, the summations of the occupancies over all projections are not the same as the total band occupancy.}
Therefore, the orbital-resolved energy contributions are approximate.

Concerning the $b_{3}$ constant, where the shape of the $ab$-basis is changed (Fig.~\ref{fig:magel_chard}c), the FM and AFM1 states prefer the magnetization axis along the elongated $a$ axis, $E_{\alpha_{1}} -  E_{\alpha_{2}}$  is negative (Table~\ref{Tab:multi_tab}). In the case of FM, it  comes from the charge density differences between Mn and Pt (Fig.~\ref{fig:magel_chard}c) -- see dominant energy contributions of $d_{xz}$ and $d_{yz}$ orbitals (Fig.~\ref{fig:EDiff_MAE}a).  For the AFM1, the situation is more complex as opposite spin appears within the layer. Nevertheless, the mechanism  is similar to that in the FM case. The orbital-resolved contributions do not help, since there are opposite contributions that almost cancel out, and the projections do not provide enough accurate information, as mentioned above. The positive sign of the AFM2 $b_{3}$ coefficients (Table~\ref{Tab:multi_tab}) can be attributed to the denoted strong  charge density difference (Fig.~\ref{fig:magel_chard}c)  from $d_{xz}$ resp. $d_{yz}$ orbitals on Mn site (Fig.~\ref{fig:EDiff_MAE}c) behaving in the opposite way to the FM state.  Therefore,  the magnetization alignment perpendicular to the elongation is preferred.

Further,  $b_{4}$ coefficient is related to the modification of the $ac$-diagonal length. The FM state prefers moments aligned along the elongated diagonal, whereas the AFM states favor the antiparallel moments pointing along the shorter one (Fig.~\ref{fig:magel_chard}d). {However, in the case of AFM1 phase, the preference is tiny.}
{The FM state favors $\alpha_{1}$ direction which comes }from $d_{xz}$ states (Fig.~\ref{fig:EDiff_MAE}a) that bring the extra charge along the $ca$ diagonal. Particularly, the Mn contributions (Fig.~\ref{fig:magel_chard}d) show slight asymmetry in charge density differences with respect to the magnetization directions. However, they are small. Concerning AFM states, $\Delta\rho$  indicates that $\alpha_{1}$ brings extra density to the shrinking Mn-Pt direction, whereas for $\alpha_{2}$ the density appears with the elongated direction, which {is} preferred. Regarding the AFM1 phase, the charge density differences near Pt atoms spread more or less near the basal plane compared to the AFM2 one (Fig.~\ref{fig:magel_chard}d). Thereby, the AFM1 $b_{4}$ magnitude is negligible, unlike the AFM2 case where  is a proximity of the Mn and Pt $\Delta\rho$ clouds.

Finally, the $b_{3}^{\prime}$ coefficients exhibit similar values irrespective of the magnetic state (Table~\ref{Tab:multi_tab}), particularly the FM and AFM2 phase. All the systems prefer spins ordered parallel  along the squeezed $ab$-diagonal with similar $b_{3}^{\prime}$ magnitude (Fig.~\ref{fig:magel_chard}e). The explanation is  simple for the FM and AFM2 phases, where the $\alpha_{2}$ axis reduces the charge density for the shrinking  Mn-Pt direction. Regarding the AFM1, the distributions of $\Delta\rho$ are quite complex, and hence it is not clear from which region the substantial interactions come. However, the orbital resolved energy contributions (Fig.~\ref{fig:EDiff_MAE}a) suggest an effect of Mn $d_{xz}$ and $d_{yz}$ orbitals  providing significant $\Delta\rho$ differences in the vicinity of the Mn atoms  (Fig.~\ref{fig:magel_chard}e).

% \FloatBarrier
%%%%%%%%########################

%\input{section_figures}
%\input{section_tables}

% \FloatBarrier

%%%%%%%%%%%%%%%%%%%%%%%%%%%%%%%%%%%%%%%%%%%%%%%%%%%%%%%%%%%%%%%%%%%%%%%%%%%%%%%%%%%
%  Conclusions
%%%%%%%%%%%%%%%%%%%%%%%%%%%%%%%%%%%%%%%%%%%%%%%%%%%%%%%%%%%%%%%%%%%%%%%%%%%%%%%%%%%

\section{Conclusions}
% \section{Conclusions}

In conclusion, we showed a significant predetermination of the magnetoelastic properties based on the type of magnetic ordering  {and explained the experimentally measured magnetostriction} in the studied  MnPt system. Due to the tetragonal symmetry, it can provide more interesting behavior than the commonly studied cubic-like materials.  With regard to the antiferromagnetic ground state, the magnetoelastic effects are much smaller than those for the ferromagnetic ordering. Possibly, the strong magnetic coupling yielding high N\'{e}el temperature makes the system insensitive to the applied field. 
However, for the ferromagnetic state observed in quenched powders and sputtered films, the magnetoelastic response is enormous, exceeding the effect in the related FePt compound~\cite{Nieves_sss_r25_seconorder}. 
We probed the origin of the differences in the magnetoelastic behavior by  analyzing charge density differences and orbital-resolved MAE contributions. It revealed a substantial  difference in  orbitals' contributions  to the magnetoelastic behavior, explaining the distinct nature of the magnetoelastic effects in the studied magnetic phases. 
The ab-initio based results with atomic spin simulations were employed  to describe observed experimental behavior, explaining well the  measured dependencies.

%\FloatBarrier

%\section*{Code availability}

\section*{Declaration of Interest Statement}
The authors declare that they have no known competing financial interests or personal relationships that could have appeared to influence the work reported in this paper.

%%%%%%%%%%%%%%%%%%%%%%%%%%%%%%%%%%%%%%%%%%%%%%%%%%%%%%%%%%%%%%%%%%%%%%%%%%%%%%%%%%%
%   Author contributions
%%%%%%%%%%%%%%%%%%%%%%%%%%%%%%%%%%%%%%%%%%%%%%%%%%%%%%%%%%%%%%%%%%%%%%%%%%%%%%%%%%%
\section*{Author contributions}

\textbf{JŠ}: Conceptualization, Data curation (ab-initio), Formal analysis, Investigation (ab-initio), Visualization, Writing – original draft;  \textbf{KS}: Data curation(experiment), Investigation (experiment), Visualization,  Writing – review \& editing;  \textbf{MV}: Data curation (experiment), Investigation (experiment), Visualization,  Writing – review \& editing; \textbf{TH}: Data curation(experiment), Investigation (experiment) ; \textbf{TJB}: Data curation(experiment), Investigation (experiment); \textbf{PN}: {Conceptualization} Writing – review \& editing; \textbf{DL}: Conceptualization, Formal analysis, Writing – review \& editing

%%%%%%%%%%%%%%%%%%%%%%%%%%%%%%%%%%%%%%%%%%%%%%%%%%%%%%%%%%%%%%%%%%%%%%%%%%%%%%%%%%%
%  Acknowledgements
%%%%%%%%%%%%%%%%%%%%%%%%%%%%%%%%%%%%%%%%%%%%%%%%%%%%%%%%%%%%%%%%%%%%%%%%%%%%%%%%%%%
\section*{Acknowledgemets}
{JS acknowledges GAČR project No. 24-11388I and DL GAČR project No. 25-14529L of the Grant Agency of Czech Republic and the CPU time by the Ministry of Education, Youth and Sports of the Czech Republic through the e-INFRA CZ (ID:90254) project.
{DL acknowledges project QM4ST (CZ.02.01.01/00/22\_008/0004572) by The Ministry of Education, Youth and Sports of the Czech Republic.}
 K.S. appreciates the support by the National Science Centre, Poland under the OPUS call in the Weave programme 2023/51/I/ST11/02562. {P. N. acknowledges support by grant MU-23-BG22/00168 funded by The Ministry of Universities of Spain.}

%%%%%%%%%%%%%%%%%%%%%%%%%%%%%%%%%%%%%%%%%%%%%%%%%%%%%%%%%%%%%%%%%%%%%%%%%%%%%%%%%%%
%%%%%%%%%%%%%%%%%%%%%%%%%%%%%%%%%%%%%%%%%%%%%%%%%%%%%%%%%%%%%%%%%%%%%%%%%%%%%%%%%%%

\FloatBarrier


\appendix

\section{Data availability}
The datasets  are available from the Zenodo repository at  https://doi.org/10.5281/zenodo.15536903

%\FloatBarrier

\section{Magnetoelasticity  - tetragonal system}

Regarding the studied MnPt system bearing the crystal symmetry of the tetragonal (I) system, the elastic energy has the following form in the Cartesian axes~\cite{AELAS_r17}, 
\begin{align}
    &\frac{1}{V_{0}}( E_{\mathrm{el}}-E_{0}) =   \label{Eq.ElasticSI}  \\
    &=\frac{1}{2}C_{11}(\varepsilon_{\mathrm{1}}^{2} + \varepsilon_{\mathrm{2}}^{2}) + C_{12}\varepsilon_{\mathrm{1}}\varepsilon_{\mathrm{2}} + C_{13}(\varepsilon_{\mathrm{1}} + \varepsilon_{\mathrm{2}}) \varepsilon_{\mathrm{3}} \nonumber \\
    &+ \frac{1}{2}C_{33} \varepsilon_{\mathrm{3}}^{2} +\frac{1}{2} C_{44}(\varepsilon_{\mathrm{4}}^{2} + \varepsilon_{\mathrm{5}}^{2}) + \frac{1}{2}C_{66} \varepsilon_{\mathrm{6}}^{2} \nonumber \\
    &=\frac{1}{2}c_{xxxx}(\varepsilon_{\mathrm{xx}}^{2} + \varepsilon_{\mathrm{yy}}^{2}) + c_{xxyy}(\varepsilon_{\mathrm{xx}}  \varepsilon_{\mathrm{yy}})  \nonumber \\
    &+ c_{xxzz}(\varepsilon_{\mathrm{xx}} + \varepsilon_{\mathrm{yy}}) \varepsilon_{\mathrm{zz}} \nonumber + \frac{1}{2}c_{zzzz} \varepsilon_{\mathrm{zz}}^{2} \\
    & +2 c_{yzyz}(\varepsilon_{\mathrm{yz}}^{2} + \varepsilon_{\mathrm{zx}}^{2}) + 2c_{xyxy} \varepsilon_{\mathrm{xy}}^{2} \nonumber \, ,
\end{align}
where $E_{0}$ and $V_{0}$ stand for the equilibrium energy and volume. {$C_{ij}$ resp. $c_{ijkl}$ denotes 6 independent elastic constants related to strain tensor matrix elements }

\begin{equation}
    \begin{pmatrix}
        \varepsilon_{\mathrm{1}} \\
        \varepsilon_{\mathrm{2}} \\
        \varepsilon_{\mathrm{3}} \\
        \varepsilon_{\mathrm{4}} \\
        \varepsilon_{\mathrm{5}} \\
        \varepsilon_{\mathrm{6}} \\    
    \end{pmatrix}
    =
    \begin{pmatrix}
        \varepsilon_{\mathrm{xx}} \\
        \varepsilon_{\mathrm{yy}}\\
        \varepsilon_{\mathrm{zz}}\\
       2 \varepsilon_{\mathrm{yz}}\\
       2 \varepsilon_{\mathrm{zx}}\\
       2 \varepsilon_{\mathrm{xy}}
    \end{pmatrix}
    \quad .
\end{equation}

On the other hand, the magnetoelastic energy possessing in the tetragonal (I) case 7 independent  magnetoelastic constants $b_{i}$ ~\cite{Callen_r65,Fritsch_r12} reads 
\begin{align}
   \frac{1}{V_{0}} E_{me} &=b_{11}(\varepsilon_{\mathrm{xx}} + \varepsilon_{\mathrm{yy}})  + b_{12}\varepsilon_{\mathrm{zz}}  \label{Eq.magel_const_SI} \\ & + b_{21} (\alpha_{z}^{2} - \frac{1}{3})(\varepsilon_{\mathrm{xx}} + \varepsilon_{\mathrm{yy}} ) +  b_{22} (\alpha_{z}^{2} - \frac{1}{3})\varepsilon_{\mathrm{zz}} \nonumber \\ & +  \frac{1}{2} b_{3} (\alpha_{x}^{2} - \alpha_{y}^{2}) (\varepsilon_{\mathrm{xx}} - \varepsilon_{\mathrm{yy}}) + 2 b_{3}^{\prime} \alpha_{x} \alpha_{y} \varepsilon_{\mathrm{xy}}  \nonumber  \\ &+ 2 b_{4} ( \alpha_{x} \alpha_{z} \varepsilon_{\mathrm{xz}} + \alpha_{y} \alpha_{z} \varepsilon_{\mathrm{yz}}) \nonumber \, .
\end{align}
The first line describes the isotropic volume effect with constants $b_{11}$ and $b_{12}$. The other lines are related to the anisotropic behavior.

Evaluating the equilibrium strain given by the minimization of the sum of elastic and magnetoelastic energy
\begin{equation}
    \frac{\partial  E_{el} + E_{me}}{\partial \varepsilon^{{eq}}_{ij}} = 0  \, ,
\end{equation}
 the relative length change  in the tetragonal (I) system follows~\cite{MAELAS_1_r21}
\begin{align}
    & \left.\frac{\Delta l }{l_{0}} \right\vert^{\bm{\alpha}}_{\bm{\beta}}  = \lambda^{\alpha 1,0} (\beta_{x}^{2} + \beta_{y}^{2}) 
    + \lambda^{\alpha 2,0} \beta_{z}^{2} \label{Eq.rel_l_change} \\  &  +  \lambda^{\alpha 1,2} (\alpha_{z}^2-\frac{1}{3}) (\beta_{x}^{2} + \beta_{y}^{2}) + \lambda^{\alpha 2,2} (\alpha_{z}^2-\frac{1}{3}) \beta_{z}^{2}
    %\nonumber \\
    \nonumber  \\
    &  +\frac{1}{2}  \lambda^{\gamma,2} (\alpha_{z}^2-\alpha_{y}^{2}) (\beta_{x}^{2}  - \beta_{y}^{2}) + 2\lambda^{\delta,2} \alpha_{x}\alpha_{y}\beta_{x}\beta_{y} \nonumber \\
    \nonumber & + 2\lambda^{\varepsilon,2} (\alpha_{x}\alpha_{z}\beta_{x}\beta_{z} + \alpha_{y}\alpha_{z}\beta_{y}\beta_{z}) \nonumber \, ,
\end{align}

%%%%%%%%%%%%%%%%%%%
 where the isotropic magnetostrictive coefficients $\lambda$ describing the volume magnetostriction are functions of the elastic muduli $C_{ij}$ and magnetoelastic constants $b_{i}$~\cite{maelas3_nieves_r23}

\begin{equation}
        \lambda^{\alpha 1,0} = \frac{-b_{11}C_{33}+b_{12}C_{13}}{C_{33}(C_{11}+C_{12})-2C_{13}^2} \, , \label{Eq.L_a10}
\end{equation}
\begin{equation}
        \lambda^{\alpha 2,0} = \frac{2b_{11}C_{13}-b_{12}(C_{11}+C_{12})}{C_{33}(C_{11}+C_{12})-2C_{13}^2} \, .\label{Eq.L_a20}
\end{equation}

Whereas the magnetization direction dependent relative length change (Eq.~\ref{Eq.rel_l_change})  is given by  anisotropic coefficients~\cite{MAELAS_1_r21} 
\begin{equation}
    %\nonumber
    \lambda^{\alpha 1,2} = \frac{-b_{21}C_{33}+b_{22}C_{13}}{C_{33}(C_{11}+C_{12})-2C_{13}^2} \label{Eq.L_a12}
\end{equation}
\begin{equation}
    %\nonumber
    \lambda^{\alpha 2,2} = \frac{2b_{21}C_{13}-b_{22}(C_{11}+C_{12})}{C_{33}(C_{11}+C_{12})-2C_{13}^2} \label{Eq.L_a22}
\end{equation}
\begin{equation}
    %\nonumber
    \lambda^{\gamma, 2} = \frac{-b_{3}}{C_{11}-C_{12}} \label{Eq.L_g2}
\end{equation}
\begin{equation}
    %\nonumber
    \lambda^{\delta, 2} = \frac{-b_{3}^{\prime}}{2C_{66}} \label{Eq.L_d2} % 
\end{equation}
\begin{equation}
    %\nonumber
    \lambda^{\varepsilon, 2} = \frac{-b_{4}}{2C_{44}} \, .
    \label{Eq.L_e2}
\end{equation}

\section{Magnetic structures}

The calculated  relaxed structure parameters are in agreement with the literature~\cite{WANG_jmmm_r13,Lu_prb_r10,Kang_prb_r23,Ravindran_PRB_01} (Table~\ref{Tab:SI1_en_str}).
Similarly, the obtained energy differences between considered magnetic phases  agree with the published ones~\cite{QURATULAIN_jmmm_r18,Lu_prb_r10} 
{, and the estimated magnitudes of spin % and \textbf{orbital
magnetic moments correspond to the literature}~\cite{Lu_prb_r10,WANG_jmmm_r13,Kang_prb_r23}.
{Alike the FePt system, the Pt sublattices bear no magnetic moments except the FM phase}~\cite{Lu_prb_r10}. {In considered AFM systems, only Mn atoms are magnetic, which can facilitate understanding of the magnetic behavior. Particularly, non-zero AFM oriented moments would break the tetragonal symmetry of the AFM1 structure in the $ab$-direction.}
%
%
%{To validate the performed calculations, experimental results were obtained. Prior to the magnetostriction measurements, the crystal structure was also explored. 
%XRD analysis indicated that the material crystallizes in a tetragonal CuAu-I type structure (space group \textit{P4/mmm}, No. 123)} (Fig.~\ref{fig:xrd-pattern}).
%
Regarding the prepared polycrystalline sample,  determined values of the lattice parameters are $a = 2.8267(2)$ Å and $c = 3.6755(3)$ Å. They are in good agreement with the literature data \cite{Andresen01061965} {and calculated AFM1 ground state, where the difference is about 1$\%$.}

\begin{table}[h]
\centering
\caption{\label{Tab:SI1_en_str}%
Calculated MnPt magnetic phase energy differences and related structure parameters.  The structure data are shown according to the FM primitive cell axes, except the $a_{nr}$ denoting the basal edge in the non-reduced cell with 4 atoms in the basis. XRD refined data of the prepared sample is added for comparison. }

\begin{tabular}{@{}c@{}cccccc@{}}
\hline
\hline
 &$\Delta$ E/f.u.& a & c & c/a & V & c/a$_{nr}$  \\
 &  (eV) & (\AA) & (\AA) &  & (\AA$^{3}$) \\
\hline

NM & 0.00 & 2.642 & 3.770 & 1.43 & 26.3 & 1.01\\
FM & -1.02 & 2.931 & 3.507 & 1.20 & 30.1 & 0.85\\
AFM1 & -1.32 & 2.799 & 3.718 & 1.33 & 29.1  & 0.94\\
AFM2  &-1.05 & 2.861 & 3.645 & 1.27 & 29.8 & 0.90\\
\hline
exp. &  & 2.827 & 3.676  & 1.30 & 29.4 &0.92 \\
\hline
\hline
\end{tabular}
\end{table}

\begin{table}[h]
\centering
\caption{\label{Tab:mag_moments}%
 MnPt spin and orbital magnetic moment components with the respect to various magnetic orderings.}
\begin{tabular}{c|ccc|ccc}
\hline
\hline
& \\[-2.5ex]
 & \multicolumn{3}{c|}{  $\mu_{\mathrm{Mn}}^{\mathrm{L}}$ ($\mu_{\mathrm{B}}$)}   & \multicolumn{3}{c}{  $\mu_{\mathrm{Pt}}^{\mathrm{L}}$ ($\mu_{\mathrm{B}}$)}  \\
 & x & y & z  & x & y & z\\
\hline 

FM & 0.00 & 0.00 & 0.03 & 0.00 & 0.00  & 0.07\\
AFM1 &  0.00 & 0.00 & 0.04& 0.00  & 0.00 & 0.00\\
AFM2  & 0.00 & 0.00 & 0.03 & 0.00  & 0.00 & 0.00\\

\hline
\hline& \\[-2.5ex]

 & \multicolumn{3}{c|}{  $\mu_{\mathrm{Mn}}^{\mathrm{S}}$ ($\mu_{\mathrm{B}}$)}   & \multicolumn{3}{c}{  $\mu_{\mathrm{Pt}}^{\mathrm{S}}$ ($\mu_{\mathrm{B}}$)}  \\
 & x & y & z  & x & y & z\\
\hline

FM & 0.00 & 0.00 & 3.82 & 0.00 & 0.00  & 0.37\\
AFM1 &  0.00 & 0.00 & 3.64 & 0.00  & 0.00 & 0.00\\
AFM2  & 0.00 & 0.00 & 3.76 & 0.00  & 0.00 & 0.00\\
\hline
\hline

\end{tabular}
\end{table}

\FloatBarrier

\section{DOS}

Density of the states (DOS) for the magnetic systems was compared (Fig.~\ref{fig:DOS}). {According to the literature, the formation of a pseudo-gap for the AFM1 ground state was observed. It is assumed that the suppressed DOS around the Fermi level, pseudogap formation, is attributed to an antiferromagnetic staggered field}~\cite{Kubota_APL_r07}. Alike behavior does not occur for the FM and AFM2 phases. Particularly, AFM2 possesses a high DOS at the  Fermi level.  It points out significant differences in the electronic structures between the AFM phases given by different orientations of the antiferromagnetic ordering. Nevertheless, for all the magnetic phases, the intensity of the Mn and Pt states is similar below the Fermi level, as the Pt conduction d-states spread across more than 6~eV while the Mn 3$d$-states tend to have similar weight as 5$d$ Pt states only from 3~eV below and up to the Fermi level.  Above, Mn dominates the conduction states for the studied energy range.

%Relation of the DOS to the exchange interactions

\begin{figure*}[h]
%(a)\hspace{-20pt}
\includegraphics[width=0.99\textwidth]{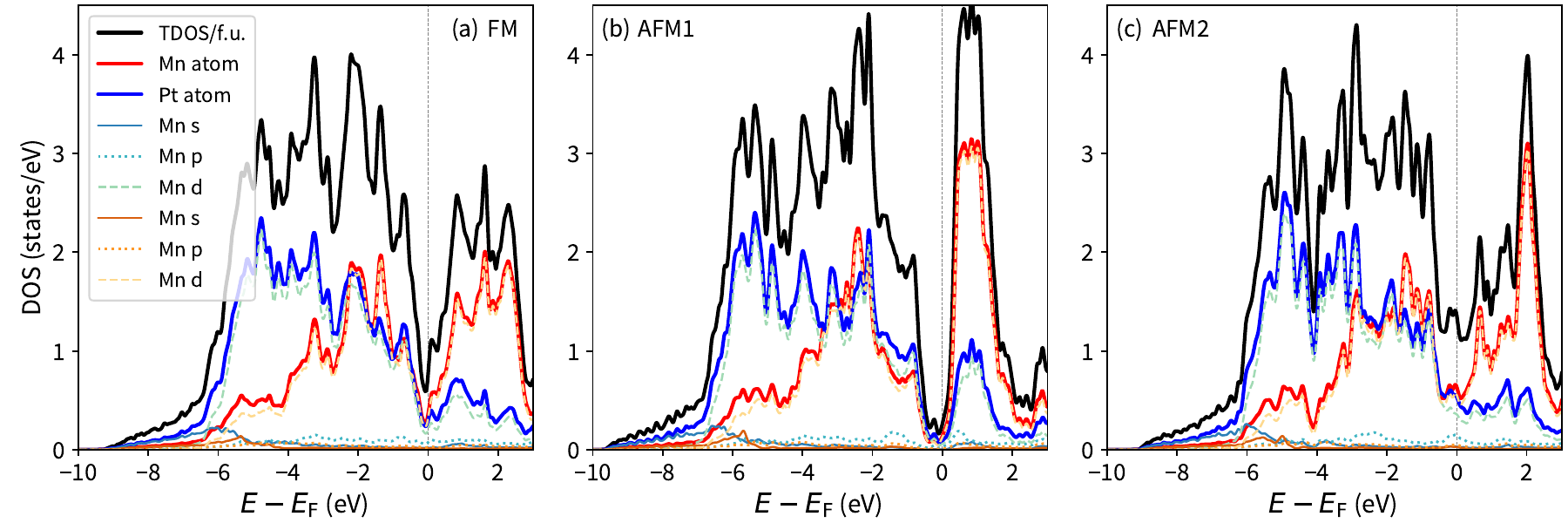}
\caption{\label{fig:DOS} Density of states with  respect to the magnetic phase. (a) FM, (b) AFM1, (c) AFM2. %Pseudogap formation in the AFM1~\cite{Kubota_APL_r07}
}
\end{figure*}

\FloatBarrier

 \section{Experimental details - XRD measurements}

The prepared polycrystalline sample was characterized by  X-ray diffraction (XRD) measurements  at room temperature employing CuK$\alpha$ radiation  ($\lambda = 1.5406$ Å) in a Bragg-Brentano geometry,  where XRD pattern was analyzed using FullProf software \cite{RODRIGUEZCARVAJAL199355}. 

% To compare the calculations with an experiment,  a polycrystalline sample of MnPt was prepared using an arc melting process with the MAM-1 system (Edmund Bühler GmbH). Stoichiometric amounts of high-purity manganese (99.9\%) and platinum (99.99\%) were melted under a titanium-gettered argon atmosphere. To ensure homogeneity, the sample was flipped and remelted multiple times. The final sample weighed approximately 1~g, with a mass loss of less than 0.5\%. No further heat treatment was applied.
% X-ray diffraction (XRD) measurements were conducted at room temperature on a sample that had been hand-ground. These measurements were performed using a PANalytical X’pert Pro diffractometer, employing CuK$\alpha$ radiation produced at 40 kV and 30 mA ($\lambda = 1.5406$ Å) in a Bragg-Brentano geometry. The resulting XRD pattern was analyzed using FullProf software \cite{RODRIGUEZCARVAJAL199355}. 
% High-resolution magnetostriction measurements, tracking length changes as a function of magnetic field (magnetostriction) at 2~K, were performed using a miniature capacitance dilatometer~\cite{Rotter_r98_magnetometer}. %
% The dilatometer was connected to an AH2500A capacitance bridge, integrated into a Physical Property Measurement System (PPMS) from Quantum Design.

\begin{figure}[h]
    \centering
    \includegraphics[width=0.8\columnwidth]{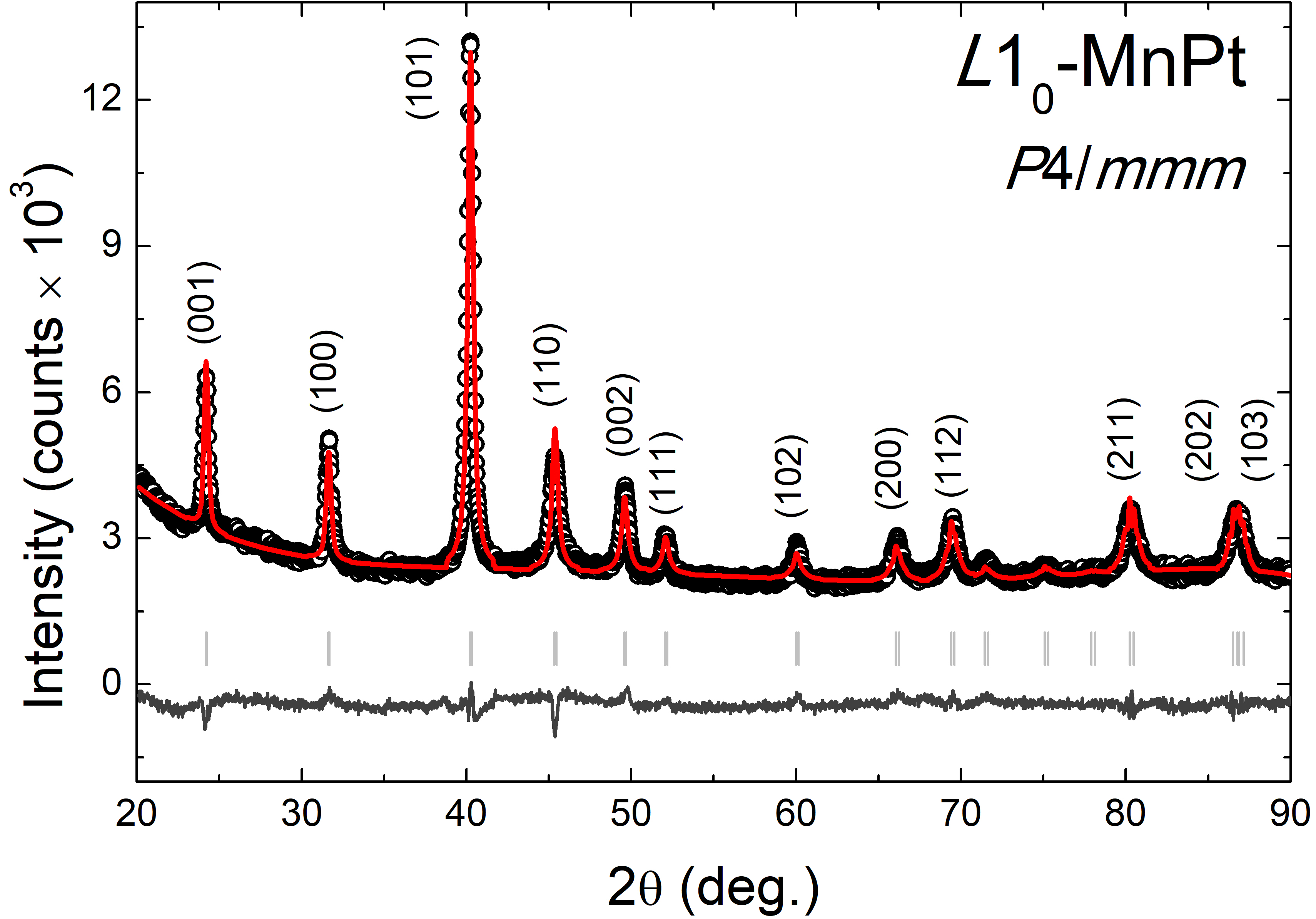}
    \caption{  The room temperature X-ray diffraction pattern for the MnPt sample.         The open circles represent the experimental data, while the solid lines depict the         Rietveld-refined pattern obtained using Fullprof software. The difference pattern is  illustrated by the solid line at the bottom. Ticks indicate the positions of the  Bragg reflections corresponding to the tetragonal CuAu-I type structure   (space group \textit{P4/mmm}, No. 123). The most prominent peaks are labeled with their Miller (\textit{hkl}) indices.
    }

    \label{fig:xrd-pattern}

\end{figure}

\FloatBarrier

\section{Elastic behavior}

The determination of magnetoelastic parameters requires knowledge of elastic coefficients  $C_{ij}$.  Since in this work we deal not only with the ground state AFM1 magnetic structure but also with the FM and AFM2, elastic coefficients were estimated in all cases, including the NM state for comparison. To be able to compare the results (Table~1)  to existing literature data for NM~\cite{WANG_jmmm_r13} and for AFM1~\cite{AISSAT_jmmm_r22}), the $C_{ij}$ are defined according to the axes of the FM primitive cell. It means a rotation of the AFM1 cell and related coefficients in the $ab$-plane by 45 degrees (Fig.~1). 
According to the calculated values, all the structures fulfilled \textit{Born stability criteria} for tetragonal (I) systems~\cite{Mouhat_prb_r14,Legut_jpcm_r12}: $C_{ii}>0$; $C_{11}>\vert C_{12}\vert$ ; $2C_{13}^{2}< C_{33} ( C_{11} + C_{12} )$.

% % tab 1
\begin{table}
\centering
\caption{\label{Tab:SI3_elast}%
The elastic constants of tetragonal MnPt for non-magnetic (NM), ferromagnetic (FM), and two antiferromagnetic orderings (AFM1 and AFM2). The last column shows the AFM1 elastic constants with respect to the axis of the non-reduced (nr) AFM1 cell, for better comparison with the published data.  % 
}
\begin{tabular}{@{}c|cccc|c@{}}
\hline
\hline

$C_{\mathrm{ij}}$(GPa) & NM & FM & AFM1  & AFM2& AFM1$^{\mathrm{nr}}$\\
\hline
$C_{\mathrm{11}}$ & 549 & 252 & 306 & 238 & 258 \\
$C_{\mathrm{12}}$ & 109 &  87 &  72 &  82 & 120\\
$C_{\mathrm{13}}$ & 171 & 152 & 141 & 155 & 141\\
$C_{\mathrm{33}}$ & 424 & 224 & 284 & 220 & 284\\
$C_{\mathrm{44}}$ & 195 &  99 & 123 &  87 & 123\\
$C_{\mathrm{66}}$ & 115 &  69 &  69 &  48 & 117\\
\hline
\hline

\end{tabular}
\end{table}

%\section{Elastic properties}
{Based on the calculated values, the NM phase has the highest bulk modulus $B$} (Table~\ref{Tab:bulk_modulus}), {making the phase the toughest. It is related to the smallest volume. The bulk modulus  of the other magnetic ones is similar, corresponding to the similar volume, particularly of FM and AFM2, possessing comparable $C_{ij}$.} 
%\DL{\texttt{Perhaps we could say more...what is the G(shear modulus) and the B/G (so-called Pugh ratio), does it differ?}}
{Regarding the shear modulus $G$, the rigidness is quite dependent on the magnetic phase, while the magnetic phases are rather ductile according to the Pugh ratio ($G/B<0.57$)}~\cite{Senkov_SciRep_r21} {unless the phase is non-magnetic, see }{(Table~\ref{Tab:bulk_modulus})}.

\begin{table}[h]
\centering
\caption{\label{Tab:bulk_modulus}%
Bulk modulus ($B$), shear modulus ($G$) and Pugh ratio ($G/B$) in the Hill approximation
}
\begin{tabular}{c|ccc}

\hline

&$B$(GPa) &$G$(GPa) & $G/B$\\
\hline

NM       &  269   &  167  &  0.624 \\
FM       &  167   &  69   &  0.414 \\
AFM1     &  178   &  95   &  0.535 \\
AFM2     &  162   &  58   &  0.355 \\

\hline

\end{tabular}
\end{table}

\FloatBarrier

\section{Magnetoelasic properties} 

{For completeness, calculated AFM1 magnetoelastic constants were estimated with respect to the axes of the non-reduced AFM1 cell }(Table~\ref{Tab:magnetostriction_AFM1}), meaning a $\pi/4$ rotation along the $c$-axis (Fig.~1b) with respect to the results mentioned in the main text (Table~1). 
Rotation of the $a$ and $b$ axes in the basal plane leads to changes in the $b_{i}$ coefficients.  The signs of $b_{21}$   and  $b_{22}$ coefficients are kept. 
%\JS{\sout{, while their magnitudes are reduced. One can relate it to the $K_{3}$ anisotropic constant and reduction of the MAE with the $\pi/4$ rotation (Fig.~\ref{fig:MAE}b).}} 
{However}, the change of the coordinates switches the signs of the $b_{3}$ and $b_{3}^{\prime}$ coefficients related to the magnetization axes in the basal plane.
{One can relate it to the $K_{3}$ anisotropic constant and reduction of the MAE with the $\pi/4$ rotation (Fig.~3b).}
Eventually, $b_{4}$ becomes negligible. Therefore, it is important to choose the proper reference basis.
In addition, the magnetoelastic coefficients were obtained not only for the tetragonal symmetry but also for the orthorhombic one (Table~\ref{Tab:magnetostriction_AFM1}). It proves that the tetragonal symmetry considerations  are valid for the AFM1 structure.

\begin{table}[h]
\centering
\caption{\label{Tab:magnetostriction_AFM1}%
 Magnetoelastic constants and magnetostrictive coefficient for AFM1 related to the non-reduced cell. Tetragonal or orthorhombic symmetry is considered.}
\begin{tabular}{crcrcrcr}
\hline
\hline

 \multicolumn{4}{c}{tetragonal} &  \multicolumn{4}{c}{orthorhombic} \\
\hline
 \multicolumn{2}{c}{$b$}   &  \multicolumn{2}{c}{$\lambda$} & \multicolumn{2}{c}{$b$}     &  \multicolumn{2}{c}{$\lambda$}   \\
  \multicolumn{2}{c}{(MPa)}  & \multicolumn{2}{c}{(10$^{-6}$)} & \multicolumn{2}{c}{(MPa)}  & \multicolumn{2}{c}{(10$^{-6}$)} \\
\hline

$b_{\mathrm{21}}$ &  9 & $\lambda^{\mathrm{\alpha 1,2}}$ & -71 & $b_{\mathrm{1}}$ & 11 & $\lambda^{\mathrm{1}}$ & -81 \\
$b_{\mathrm{22}}$ &  -15 & $\lambda^{\mathrm{\alpha 2,2}}$ & 126 & $b_{\mathrm{2}}$ & -33 & $\lambda^{\mathrm{2}}$ & 232 \\
$b_{\mathrm{3}}$ &   43 & $\lambda^{\mathrm{\gamma, 2}}$ & -310& $b_{\mathrm{3}}$ & -33 & $\lambda^{\mathrm{3}}$ & 230\\
$b_{\mathrm{4}}$ &   1 & $\lambda^{\mathrm{\varepsilon, 2}}$ & -2& $b_{\mathrm{4}}$ & 11 & $\lambda^{\mathrm{4}}$ & -82\\
$b_{\mathrm{3}}^{\prime}$ & -37 & $\lambda^{\mathrm{\delta, 2 }}$ & 158 & $b_{\mathrm{5}}$ & 14 & $\lambda^{\mathrm{5}}$ & -122 \\
& & &  &$b_{\mathrm{6}}$ & 13 & $\lambda^{\mathrm{6}}$ & -121  \\
& & &  &$b_{\mathrm{7}}$ & -37 & $\lambda^{\mathrm{7}}$ &  154 \\
& & &  &$b_{\mathrm{8}}$ & 1 & $\lambda^{\mathrm{8}}$ &  -52 \\
& & &  &$b_{\mathrm{9}}$ & 1 & $\lambda^{\mathrm{9}}$ & -52 \\
\hline
\hline

\end{tabular}
\end{table}

\section{Magnetoelasticity - orthorhombic system}

{Regarding an orthorhombic system, the magnetoelastic energy is described by 12 independent magnetoelastic constants (3 isotropic and 9 anisotropic ones) as follows~\cite{MAELAS_1_r21}}
\begin{align}
   \frac{1}{V_{0}} E_{me} &=b_{01}\varepsilon_{\mathrm{xx}}  + b_{02}\varepsilon_{\mathrm{yy}} + b_{03}\varepsilon_{\mathrm{zz}}   \\
   & + b_{1} \alpha_{x}^{2}\varepsilon_{\mathrm{xx}} + b_{2} \alpha_{y}^{2}\varepsilon_{\mathrm{xx}} +  b_{3} \alpha_{x}^{2}\varepsilon_{\mathrm{yy}} \nonumber \\
      & + b_{4} \alpha_{y}^{2}\varepsilon_{\mathrm{yy}} + b_{5} \alpha_{x}^{2}\varepsilon_{\mathrm{zz}} +  b_{6} \alpha_{y}^{2}\varepsilon_{\mathrm{zz}}  \nonumber\\
   &+ 2 b_{7} \alpha_{x}\alpha_{y}\varepsilon_{\mathrm{xy}} + 2 b_{8} \alpha_{x}\alpha_{z}\varepsilon_{\mathrm{xz}} + 2 b_{9} \alpha_{y}\alpha_{z}\varepsilon_{\mathrm{yz}} \nonumber  \, .
\end{align}

{The relative length change at the equilibrium strain in the orthorhombic system follows~\cite{MAELAS_1_r21}}
\begin{align}
    & \left.\frac{\Delta l }{l_{0}} \right\vert^{\bm{\alpha}}_{\bm{\beta}}  = \lambda^{\alpha 1,0} \beta_{x}^{2} +  \lambda^{\alpha 2,0} \beta_{y}^{2} +  \lambda^{\alpha 3,0} \beta_{z}^{2}  \\
    &  +  \lambda^{1}(\alpha_{x}^2 \beta_{x}^{2} - \alpha_{x}\alpha_{y} \beta_{x}\beta_{y} - \alpha_{x}\alpha_{z} \beta_{x}\beta_{z})
    \nonumber  \\
    &  +  \lambda^{2}(\alpha_{y}^2 \beta_{x}^{2} - \alpha_{x}\alpha_{y} \beta_{x}\beta_{y} )
    %\nonumber  \\      &
    +  \lambda^{3}(\alpha_{x}^2 \beta_{y}^{2} - \alpha_{x}\alpha_{y} \beta_{x}\beta_{y} )
    \nonumber  \\
    &  +  \lambda^{4}(\alpha_{y}^2 \beta_{y}^{2} - \alpha_{x}\alpha_{y} \beta_{x}\beta_{y} - \alpha_{y}\alpha_{z} \beta_{y}\beta_{z})
    \nonumber  \\
    &  +  \lambda^{5}(\alpha_{x}^2 \beta_{z}^{2} -\alpha_{x}\alpha_{z} \beta_{x}\beta_{z} )
    %\nonumber  \\    %     &
    +  \lambda^{6}(\alpha_{y}^2 \beta_{z}^{2} - \alpha_{y}\alpha_{z} \beta_{y}\beta_{z} )
    \nonumber  \\
    &  +  4\lambda^{7}\alpha_{x}\alpha_{y}\beta_{x}\beta_{y}
    +4\lambda^{8}\alpha_{x}\alpha_{z}\beta_{x}\beta_{z}
    +4\lambda^{8}\alpha_{y}\alpha_{z}\beta_{y}\beta_{z}
    \nonumber  
  \, .
\end{align}

\FloatBarrier

%\newpage

\section{Exchange interactions} 
The atomistic spin simulations (main text Figure 5)  employed the following isotropic exchange interaction constants up to the 6$^{th}$ nearest neighbor shell (Fig~\ref{fig:Jij_volume-dep}). The calculated magnitudes of the exchange interactions  agree with the literature~\cite{Kang_prb_r23}.

\begin{figure}
    \centering
    \includegraphics[width=0.95\columnwidth]{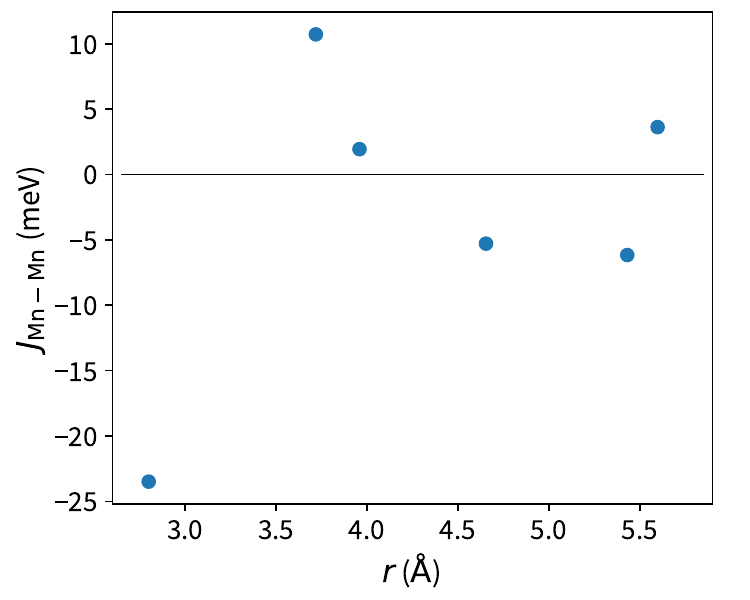}
    \caption{ Magnetic exchange interactions of the AFM1 MnPt  phase.  }
    \label{fig:Jij_volume-dep}
\end{figure}

\FloatBarrier

%\FloatBarrier

  \bibliographystyle{elsarticle-num} 
  \bibliography{mnptsamp.bib}



%\begin{thebibliography}{00}


%\bibitem[Lamport(1994)]{lamport94}
%  \textit{\LaTeX: a document preparation system},
%  Addison Wesley, Massachusetts,

%\end{thebibliography}

\end{document}